\documentclass[a4paper]{aa}
\usepackage[varg]{txfonts}

\usepackage{graphicx}
\usepackage{natbib}
\bibpunct{(}{)}{;}{a}{}{,} 
\usepackage{amsmath}
\usepackage{color}
\usepackage[breaklinks=true]{hyperref}

\newcommand{\snia}{SN~Ia}
\newcommand{\sneia}{SNe~Ia}
\newcommand{\snf}{SNfactory}
\newcommand{\snifs}{\textsc{snifs}}
\newcommand{\synapps}{\texttt{SYNAPPS}}

\newcommand{\wl}{$\lambda$}
\newcommand{\cf}{\emph{cf.}}
\newcommand{\phn}{\phantom{0}}
\newcommand{\nodata}{\ldots}

\begin{document}

\title{Spectrophotometric time series of SN 2011fe\\
 from the Nearby Supernova Factory}

\author
{
  {R.~Pereira}\inst{\ref{ipnl}}
  \and {R.~C. Thomas}\inst{\ref{berkccc}}
  \and {G.~Aldering}\inst{\ref{lbnl}}
  \and {P.~Antilogus}\inst{\ref{lpnhe}}
  \and {C.~Baltay}\inst{\ref{yale}}
  \and {S.~Benitez-Herrera}\inst{\ref{mpa}}
  \and {S.~Bongard}\inst{\ref{lpnhe}}
  \and {C.~Buton}\inst{\ref{bonn}}
  \and {A.~Canto}\inst{\ref{lpnhe}}
  \and {F.~Cellier-Holzem}\inst{\ref{lpnhe}}
  \and {J.~Chen}\inst{\ref{tsinghua}}
  \and {M.~Childress}\inst{\ref{lbnl},\ref{berkphys}}
  \and {N.~Chotard}\inst{\ref{tsinghua},\ref{nao}}
  \and {Y.~Copin}\inst{\ref{ipnl}}
  \and {H.~K. Fakhouri}\inst{\ref{lbnl},\ref{berkphys}}
  \and {M.~Fink}\inst{\ref{mpa}}
  \and {D.~Fouchez}\inst{\ref{cppm}}
  \and {E.~Gangler}\inst{\ref{ipnl}}
  \and {J.~Guy}\inst{\ref{lpnhe}}
  \and {W.~Hillebrandt}\inst{\ref{mpa}}
  \and {E.~Y. Hsiao}\inst{\ref{lbnl}}
  \and {M.~Kerschhaggl}\inst{\ref{bonn}}
  \and {M.~Kowalski}\inst{\ref{bonn}}
  \and \\{M.~Kromer}\inst{\ref{mpa}}
  \and {J.~Nordin}\inst{\ref{lbnl},\ref{ssl}}
  \and {P.~Nugent}\inst{\ref{berkccc}}
  \and {K.~Paech}\inst{\ref{bonn}}
  \and {R.~Pain}\inst{\ref{lpnhe}}
  \and {E.~P\'econtal}\inst{\ref{cral}}
  \and {S.~Perlmutter}\inst{\ref{lbnl},\ref{berkphys}}
  \and \\{D.~Rabinowitz}\inst{\ref{yale}}
  \and {M.~Rigault}\inst{\ref{ipnl}}
  \and {K.~Runge}\inst{\ref{lbnl}}
  \and {C.~Saunders}\inst{\ref{lbnl}}
  \and {G.~Smadja}\inst{\ref{ipnl}}
  \and \\{C.~Tao}\inst{\ref{tsinghua},\ref{cppm}}
  \and {S.~Taubenberger}\inst{\ref{mpa}}
  \and {A.~Tilquin}\inst{\ref{cppm}}
  \and {C.~Wu}\inst{\ref{lpnhe},\ref{nao}}\\
  (The Nearby Supernova Factory)
}

\institute{
  Universit\'e de Lyon, 69622, France; Universit\'e de Lyon 1, France;
  CNRS/IN2P3, Institut de Physique Nucl\'eaire de Lyon,
  France
  \email{rui.pereira@in2p3.fr}\label{ipnl}
  \and
  Computational Cosmology Center, Computational Research Division,
  Lawrence Berkeley National Laboratory, 1 Cyclotron Road MS~50B-4206,
  Berkeley, CA, 94611, USA\label{berkccc}
  \and
  Physics Division, Lawrence Berkeley National Laboratory,
  1 Cyclotron Road, Berkeley, CA 94720, USA\label{lbnl}
  \and
  Laboratoire de Physique Nucl\'eaire et des Hautes \'Energies,
  Universit\'e Pierre et Marie Curie Paris 6, Universit\'e Paris
  Diderot Paris 7, CNRS-IN2P3, 4 place Jussieu, 75252 Paris Cedex 05,
  France\label{lpnhe}
  \and
  Department of Physics, Yale University,
  New Haven, CT 06520-8121, USA\label{yale}
  \and
  Max-Planck-Institut f\"ur Astrophysik, Karl-Schwarzschild-Str. 1,
  85741 Garching bei M\"unchen, Germany\label{mpa}
  \and
  Physikalisches Institut, Universit\"at Bonn,
  Nu\ss allee 12, 53115 Bonn, Germany\label{bonn}
  \and
  Tsinghua Center for Astrophysics, Tsinghua University, Beijing
  100084, China\label{tsinghua}
  \and
  Department of Physics, University of California Berkeley,
  366 LeConte Hall MC 7300, Berkeley, CA, 94720-7300, USA\label{berkphys}
  \and
  National Astronomical Observatories, Chinese Academy of Sciences,
  Beijing 100012, China\label{nao}
  \and
  Centre de Physique des Particules de Marseille, Aix-Marseille
  Universit\'e, CNRS/IN2P3, 163, avenue de Luminy - Case 902 - 13288
  Marseille Cedex 09, France\label{cppm}
  \and
  Space Sciences Laboratory, University of California Berkeley, 
  7 Gauss Way, Berkeley, CA 94720, USA\label{ssl}
  \and
  Centre de Recherche Astronomique de Lyon, Universit\'e Lyon 1,
  9 Avenue Charles Andr\'e, 69561 Saint Genis Laval Cedex, France\label{cral}
}

\date{Received 26 December 2012 / Accepted 21 January 2013}


\abstract{ We present 32 epochs of optical (3300--9700~\AA)
  spectrophotometric observations of the nearby quintessential
  ``normal'' type Ia supernova (\snia) SN~2011fe in the galaxy M101,
  extending from $-\,15$ to $+\,97$~d with respect to $B$-band
  maximum, obtained by the Nearby Supernova Factory collaboration.
  SN~2011fe is the closest ($\mu=29.04$) and brightest
  ($B_\mathrm{max}=9.94$~mag) \snia\ observed since the advent of
  modern large scale programs for the intensive periodic followup of
  supernovae. Both synthetic light curve measurements and spectral
  feature analysis attest to the normality of SN~2011fe.  There is
  very little evidence for reddening in its host galaxy.  The
  homogeneous calibration, intensive time sampling, and high
  signal-to-noise ratio of the data set make it unique. Thus it is
  ideal for studying the physics of \snia\ explosions in detail, and
  for furthering the use of \sneia\ as standardizable candles for
  cosmology.  Several such applications are shown, from the creation
  of a bolometric light curve and measurement of the
  $^{56}\mathrm{Ni}$ mass, to the simulation of detection thresholds
  for unburned carbon, direct comparisons with other \sneia, and
  existing spectral templates.  }
\titlerunning{Spectrophotometric time series of SN~2011fe}
\authorrunning{R.~Pereira \& the \snf}
\maketitle


\section{Introduction}

Having exploded in the Pinwheel Galaxy just 6.4~Mpc distant
\citep[$z=0.00080\pm0.00001$; $\mu=29.04\pm0.19$,][]{Paturel03, Shappee11}, 
the Type~Ia supernova SN~2011fe represents a rare opportunity for
intensive study.  Its discovery \citep{ATEL3581} by the Palomar
Transient Factory \citep[PTF;][]{Law09, Rau09} less than 12 hours after
outburst \citep{Nugent11} precipitated a number of diverse ground-based
and space-based follow-up campaigns.  Promptly initiated, high-cadence
observations from a number of these have been published already.
\citet{Brown12} presented two months of nearly continuous
\emph{Swift}/UVOT follow-up in ultraviolet $uvw2$, $uvm2$, and $uvw1$
filters.  \emph{BVRI} photometry obtained over six months appeared in
\citet{Richmond12}, \citet{Vinko12} and \citet{Munari13}.
\citet{Tammann11} used extensive optical photometry of SN~2011fe
obtained by the American Association of Variable Star Observers
(AAVSO) and the tip of the red-giant branch distance to its host
(M101) to measure $H_0$.  Covering the near-infrared,
\citet{Matheson12} presented high-cadence \emph{JHK} photometry from the
Wisconsin Indiana Yale NOAO (WIYN) telescope with WHIRC.  The apparent
brightness ($V\sim10$ at peak) of SN~2011fe made multi-epoch
spectropolarimetry much more accessible than usual \citep{Smith11}.
Finally, \citet{Parrent12} have published 18 optical spectra from
multiple telescopes, starting 1.2 days after explosion with a 1.8 day
average cadence, along with a spectroscopic analysis including
constraints on unburned carbon.

These data sets and others can address numerous long-standing
questions about the nature of the Type~Ia supernova (\snia) progenitor
systems, environs, and explosion mechanisms.  It has generally been
thought that \sneia\ arise from the thermonuclear disruption of white
dwarf stars accreting material from a companion donor star \citep[for
an overview, see e.g.][]{Branch95}.  Using images obtained just hours
after outburst, pre-explosion X-ray limits, and the inferred $^{56}$Ni
yield, \citet{Bloom12} constrain the progenitor primary of SN~2011fe
to be a white dwarf or neutron star -- the first direct imaging
evidence for a compact primary.  Pre-explosion multi-wavelength
archival images of the SN~2011fe stellar neighborhood require the
mass-donating secondary to be another white dwarf, subgiant, or
main-sequence star; red giants and helium stars are excluded with the
companion restricted to $M < 3.5\,\mathrm{M}_\odot$ \citep{Li11}.
Radio and x-ray observations probing the circumstellar environment of
SN~2011fe imply a progenitor system with a mass loss rate as low as $6
\times 10^{-10}\,\mathrm{M}_\odot\,\mathrm{yr}^{-1}$, severely
constraining models where the secondary is not a white dwarf
\citep[]{Chomiuk12, Horesh12, Margutti12}.  A case study in attempting
to constrain explosion models from spectrophotometric observations
slightly favors a white dwarf companion \citep{Ropke12} but further
detailed modeling is needed to be more conclusive.  The future
emergence (or absence) of a surviving non-compact secondary,
brightened as a consequence of its interaction with the SN ejecta, may
be decisive in identifying the progenitor system \citep{Shappee12}.

Such progress is exciting for several reasons; chief among these is the
potential ability for such results to reduce uncertainty about the
reliability of \sneia\ as tools for observational cosmology.  The use of
\sneia\ as standardizable candles \citep{Phillips93} brought about the
discovery of the accelerating expansion of the Universe just over a
decade ago \citep{Riess98, Perlmutter99}.  \sneia\ have since become a
key means for constraining the unknown physics of cosmic acceleration,
called ``Dark Energy'' \citep[e.g.,][]{Guy10, Howell11, Suzuki12}.  A
fundamental physics result is thus inextricably linked to the
details of binary stellar evolution and stellar death, details which
SN~2011fe may help sort out.  It is thus fortuitous that SN~2011fe is
also a normal \snia, and is highly representative of the typical \snia\
sought for placement on a Hubble diagram.

In this article, the Nearby Supernova Factory
\citep[\snf,][]{Aldering02} presents an atlas of 32 spectrophotometric
observations of SN~2011fe extending from $-\,15$ to $+\,97$~d with respect
to the time of maximum light.  Some of these spectra have appeared in
the \citet{Ropke12} study and a quick-pipeline reduction of the first
spectrum presented here has appeared in \citet{Parrent12}.  This atlas
should become a useful resource in studying \snia\ physics, in
exploring systematics in the analysis of \snia\ spectra and spectral
indicators, and in constructing spectral templates for SN cosmology
applications.

The remainder of this article is organized as follows.  In  Sect.~\ref{sec:spectrophot-observ} 
we describe the observations and data reduction
procedure, present the reduced spectrophotometric time series, synthesize
photometry from it, and analyze the synthetic light curves using a standard
\snia\ light curve fitter.  Analyses of light curve residuals, possible
extinction in the host galaxy, and the bolometric light curve appear in
Sect.~\ref{sec:analysis}.  This section also covers spectral feature
measurements, sub-classification within existing schemes, parameterized
spectral fitting, and examination of unburned carbon signatures.  In
Sect.~\ref{sec:discussion}, we place our data and analysis in context but
primarily seek to demonstrate some useful features and applications of the
data set.  Conclusions appear in Sect.~\ref{sec:conclusion}.  The
spectrophotometry of SN~2011fe is available for download in electronic form at
the SNfactory project website\footnote{\url{http://snfactory.lbl.gov}} and at
the CDS via anonymous ftp to cdsarc.u-strasbg.fr (130.79.128.5) or via
\url{http://cdsweb.u-strasbg.fr/cgi-bin/qcat?J/A+A/}.


\section{Spectrophotometric observations}\label{sec:spectrophot-observ}

\begin{table*}
  \caption{Observing log for \snifs\ spectra of SN~2011fe}
  \label{tab:obslog}
  \centering
  {\footnotesize
  \begin{tabular}{ccccccccc}
    \hline\hline
    $t-t_{expl}$\tablefootmark{a} &
    $t-t_{max}$\tablefootmark{b} &
    UTC Date &
    MJD\tablefootmark{c} &
    Photometricity\tablefootmark{d} &
    Exp. Time (s) &
    Airmass &
    Seeing (\arcsec) \\
    \hline\\[-2ex]
    \phn\phn2.6 & $-15.2$ & 2011 Aug. 26.3 & 55799.3 & $\bullet$ (4) & 2 $\times$ 300 & 1.87 & 2.24\\
    \phn\phn3.5 & $-14.3$ & 2011 Aug. 27.2 & 55800.2 & $\bullet$ (1) & 2 $\times$ 300 & 1.74 & 1.04\\
    \phn\phn4.5 & $-13.3$ & 2011 Aug. 28.2 & 55801.2 & $\bullet$ (3) & 2 $\times$ 300 & 1.76 & 1.29\\
    \phn\phn5.6 & $-12.2$ & 2011 Aug. 29.3 & 55802.3 & $\bullet$ (1) & 2 $\times$ 250 & 1.87 & 1.01\\
    \phn\phn6.5 & $-11.3$ & 2011 Aug. 30.2 & 55803.2 & $\bullet$ (1) & 2 $\times$ 250 & 1.82 & 0.98\\
    \phn\phn7.5 & $-10.3$ & 2011 Aug. 31.2 & 55804.2 & $\bullet$ (3) & 2 $\times$ 250 & 1.81 & 0.94\\
    \phn\phn8.5 & \phn$-9.3$ & 2011 Sep. 01.2 & 55805.2 & $\bullet$ (1) & 2 $\times$ 250 & 1.88 & 1.09\\
    \phn\phn9.5 & \phn$-8.3$ & 2011 Sep. 02.2 & 55806.2 & $\bullet$ (2) & 2 $\times$ 250 & 1.83 & 1.95\\
    \phn10.6 & \phn$-7.2$ & 2011 Sep. 03.3 & 55807.3 & $\circ$ (1) & 2 $\times$ 250 & 2.07 & 1.88\\
    \phn11.5 & \phn$-6.3$ & 2011 Sep. 04.2 & 55808.2 & $\bullet$ (4) & 2 $\times$ 250 & 1.87 & 2.07\\
    \phn12.5 & \phn$-5.3$ & 2011 Sep. 05.2 & 55809.2 & $\bullet$ (1) & 2 $\times$ 250 & 1.88 & 1.21\\
    \phn16.5 & \phn$-1.3$ & 2011 Sep. 09.2 & 55813.2 & $\bullet$ (3) & 2 $\times$ 250 & 2.08 & 1.53\\
    \phn17.5 & \phn$-0.3$ & 2011 Sep. 10.2 & 55814.2 & $\bullet$ (3) & 2 $\times$ 250 & 1.94 & 0.99\\
    \phn18.5 & \phn\phn0.7 & 2011 Sep. 11.2 & 55815.2 & $\bullet$ (3) & 2 $\times$ 250 & 1.95 & 1.15\\
    \phn19.5 & \phn\phn1.7 & 2011 Sep. 12.2 & 55816.2 & $\bullet$ (4) & 2 $\times$ 250 & 1.99 & 1.24\\
    \phn20.5 & \phn\phn2.7 & 2011 Sep. 13.2 & 55817.2 & $\bullet$ (4) & 2 $\times$ 250 & 1.98 & 0.96\\
    \phn21.5 & \phn\phn3.7 & 2011 Sep. 14.2 & 55818.2 & $\bullet$ (1) & 2 $\times$ 250 & 2.17 & 1.94\\
    \phn24.5 & \phn\phn6.7 & 2011 Sep. 17.2 & 55821.2 & $\circ$ (3) & 2 $\times$ 300 & 2.18 & 1.01\\
    \phn26.5 & \phn\phn8.7 & 2011 Sep. 19.2 & 55823.2 & $\circ$ (2) & 2 $\times$ 250 & 2.18 & 1.56\\
    \phn29.5 & \phn11.7 & 2011 Sep. 22.2 & 55826.2 & $\circ$ (1) & 2 $\times$ 250 & 2.20 & 1.21\\
    \phn31.5 & \phn13.7 & 2011 Sep. 24.2 & 55828.2 & $\bullet$ (2) & 2 $\times$ 300 & 2.34 & 1.06\\
    \phn34.5 & \phn16.7 & 2011 Sep. 27.2 & 55831.2 & $\circ$ (2) & 2 $\times$ 250 & 2.36 & 1.19\\
    \phn36.5 & \phn18.7 & 2011 Sep. 29.2 & 55833.2 & $\circ$ (3) & 2 $\times$ 250 & 2.60 & 1.37\\
    \phn39.5 & \phn21.7 & 2011 Oct. 02.2 & 55836.2 & $\circ$ (1) & 2 $\times$ 250 & 2.75 & 1.51\\
    \phn41.5 & \phn23.7 & 2011 Oct. 04.2 & 55838.2 & $\circ$ (1) & 2 $\times$ 350 & 2.79 & 1.68\\
    \phn91.9 & \phn74.1 & 2011 Nov. 23.6 & 55888.6 & $\circ$ (4) & 1 $\times$ 250 & 2.25 & 1.29\\
    \phn94.9 & \phn77.1 & 2011 Nov. 26.6 & 55891.6 & $\circ$ (3) & 1 $\times$ 250 & 2.04 & 1.08\\
    \phn96.9 & \phn79.1 & 2011 Nov. 28.6 & 55893.6 & $\circ$ (3) & 2 $\times$ 250 & 2.05 & 1.91\\
    \phn99.9 & \phn82.1 & 2011 Dec. 01.6 & 55896.6 & $\circ$ (4) & 3 $\times$ 250 & 2.10 & 1.35\\
    104.9 & \phn87.1 & 2011 Dec. 06.6 & 55901.6 & $\circ$ (6) & 3 $\times$ 300 & 2.16 & 0.93\\
    106.9 & \phn89.1 & 2011 Dec. 08.6 & 55903.6 & $\circ$ (6) & 3 $\times$ 300 & 1.91 & 1.24\\
    114.9 & \phn97.1 & 2011 Dec. 16.6 & 55911.6 & $\circ$ (7) & 2 $\times$ 300 & 1.71 & 1.38\\
    \hline
  \end{tabular}}
  \tablefoot{In the case of multiple consecutive observations in a
    single night, the phases, dates and airmass correspond to the
    middle of the first exposure, while the seeing is the average value.\\
    \tablefoottext{a}{Days relative to the date of explosion derived by
      \citet{Nugent11}: MJD~55796.696.}\\
    \tablefoottext{b}{Phase, observer-frame days relative to $B$-band maximum
      light: MJD 55814.51.}\\
    \tablefoottext{c}{JD - 2400000.5}\\
    \tablefoottext{d}{$\bullet$ photometric, $\circ$ non-photometric, (\#)
      number of standard stars observed during the night and used for
      atmospheric extinction and telluric absorption correction.}}
\end{table*}

\subsection{Data acquisition}

The data were obtained using the SuperNova Integral Field Spectrograph
\citep[\snifs,][]{Lantz04}. \snifs\ is a fully integrated instrument
optimized for automated observation of point sources on a structured
background over the full ground-based optical window at moderate
spectral resolution. It consists of a high-throughput wide-band
pure-lenslet integral field spectrograph \citep[IFS, ``\`a la
TIGER;''][]{Bacon95,Bacon01}, a multi-filter photometric channel
to image the stars in the vicinity of the IFS field-of-view (FOV) to
monitor atmospheric transmission during spectroscopic exposures, and an
acquisition/guiding channel. The IFS possesses a fully-filled $6\farcs 4
\times 6\farcs 4$ spectroscopic field of view subdivided into a grid of
$15 \times 15$ spatial elements, a dual-channel spectrograph covering
3200--5200~\AA\ and 5100--10000~\AA\ simultaneously with FWHM
resolutions of 5.65~\AA\ and 7.54~\AA\ respectively, and an internal
calibration unit (continuum and arc lamps). \snifs\ is continuously
mounted on the south bent Cassegrain port of the University of Hawaii
2.2~m telescope (UH~2.2m) on Mauna Kea. The telescope and instrument,
under script control, are supervised remotely.

PTF discovered SN~2011fe (PTF11kly) in images obtained on 2011
Aug.~24.2 (UTC, used throughout).  \citet{Nugent11} derive an
explosion date of Aug.~23.7.  \snf\ follow-up observations commenced
on Aug.~26.3, an estimated 2.6 days after the explosion, and encompass
32 nights with two consecutive spectrophotometric exposures on most of
the nights. The observing log is shown in Table~\ref{tab:obslog}.
Daily observational cadence was maintained until 12.5 days after
explosion.  Lack of telescope access then forced a four day gap, but
the daily cadence was re-established for days 16.5 to 21.5 after
explosion. The cadence was relaxed to every 2-3 days until SN~2011fe
was no longer observable from Hawaii, 41.5 days after explosion.
Follow-up resumed a month and a half later, when observations were
made over 7 nights spanning 20 days. The last spectra reported here
were obtained Dec.~16.6, about 115 days after explosion.  The position
of SN~2011fe on the sky during the follow-up campaign forced
observations at low altitude.  The mean airmass was 2.1 and most of
the early observations were taken during astronomical twilight.

\subsection{Data calibration}\label{sec:data-calibration}

All spectra were reduced using \snf's dedicated data reduction
pipeline, similar to that presented in Sect.~4 of \citet{Bacon01}. A
brief discussion of the software pipeline is presented in
\citet{Aldering06} and is updated in Sect.~2.1 of~\citet{Scalzo10}.
SN~2011fe is an unusual target for \snifs\ and its associated data
reduction software pipeline.  The brightness of the SN itself and the
small amount of time that the target was above the horizon imposed
constraints on integration time.  This affects our ability to use our
standard multi-filter photometric channel for flux calibration on
non-photometric nights.  Also, some observations were obtained using
interrupt time limited to $\sim 1$~h which did not allow for execution
of more than one standard star observation needed for optimal
calibration.  Together, these factors limit the accuracy of our flux
calibration of SN~2011fe, despite the brightness of the target.  We
will thus elaborate on the data reduction discussion given in
\citet{Scalzo10} and on modifications needed in order to flux
calibrate the observations presented here.

Following standard low-level processing, CCD frames from both \snifs\
channels are mapped into $(x, y, \lambda)$ datacubes.  Spectra are
extracted from these using a chromatic spatial point-spread function
(PSF) assuming spatially flat background, thus removing most of the
host galaxy light as part of the sky.  Host galaxy light contamination
was evaluated using a background-subtracted stack of $5\times60$~s
CFH12k $R$-band exposures of the host galaxy of SN~2011fe (M101),
which was photometrically calibrated using SDSS $r$-band observations
of the same region. Several \snifs\ acquisition images were stacked
and astrometrically aligned to the CFH12k stack in order to determine
the location of SN~2011fe within the CFHT image. The mean $r$-band
host galaxy surface brightness, inside the \snifs\ FOV centered around
the SN, equals 22.15 mag~arcsec$^{-2}$ and is almost flat, with structured
residuals in the order of $r\sim23$ mag inside 1\arcsec\ diameter
apertures.  The remaining background structural effects are thus
considered negligible with respect to the brightness of SN~2011fe, so
the detailed host galaxy subtraction method described by
\citet{Bongard11} was not necessary.

The extracted spectra are next merged and truncated to the final
wavelength range of 3300--9700~\AA.  Flux calibration is performed
using an instrumental flux solution derived from all
spectrophotometric standard stars observed during the same night, and
a mean atmospheric extinction computed from 6 years of \snifs\ data
\citep{Buton13}. Observations of multiple standard stars were not
always possible, particularly on interrupt nights.  This limits a
proper nightly extinction computation, so for the sake of consistency
we use the mean extinction law on all nights. In this particular
framework, the precision of the flux calibration depends on the number
of standard stars observed per night, but not on their airmass
distribution.

During non-photometric nights, any achromatic (grey) differential
atmospheric attenuation between the observations of SN~2011fe and the
standard stars is accounted for using simultaneous observations of
field stars, through the \snifs\ multi-filter photometric
channel. Because SN~2011fe was very bright, relatively short exposure
times were used to avoid saturation of the spectrograph, so the number
of field stars visible at all epochs under very different attenuation
conditions is small.  After detection of the objects in the
multi-filter field, the object catalogs were manually inspected to insure proper star
selection and to prevent inaccurate inter-epoch astrometric and
photometric alignments, which are performed using a modified version
of the SuperNova Legacy Survey photometry code
\citep[\texttt{poloka},][]{Astier06}.  The flux ratio between each
exposure and a reference exposure (with the best seeing) equals the
integral of the convolution kernel needed to degrade the reference
image PSF into the ``photometric frame'' of the other image.
This night-to-night variation in the brightness of the field stars,
normalized by observations on photometric nights, provides the
photometric scaling between fields and is used as a per-exposure grey
flux correction to the mean atmospheric extinction, thus allowing
absolute flux calibration.  Seeing and night photometricity were
assessed using quantitative analyses of \snifs\ guider video frames
acquired during our exposures, along with deglitched CFHT Skyprobe
data \citep{Cuillandre02}, attenuation estimates from the instrumental
flux solution and photometric ratio computations \citep[\cf\ Sect.~5
of][]{Buton13}, and infrared satellite imagery.

The flux-calibrated spectra were corrected for telluric absorption
using a nightly average spectrum determined using standard star
observations.  While consistent with the errors, this correction is
nevertheless imperfect, and some (very small) glitches are visible in
the spectra, especially for later phases at $\lambda>9000$~\AA. Milky
Way dust extinction along the line of sight is corrected assuming
E$(B-V)=0.0088$~mag \citep{Schlegel98} and an extinction law with
$R_V=3.1$ \citep{Cardelli89,ODonnell94}.  No extinction correction for
dust in the host galaxy is applied (see Sect.~\ref{sec:dust}).
Multiple spectra obtained the same night were merged using a
variance-weighted mean.

\subsection{Spectral time series and synthesized photometry}
\label{sec:time-series}

\begin{figure*}
    \centering
    \includegraphics[width=0.9\textwidth,clip=true]{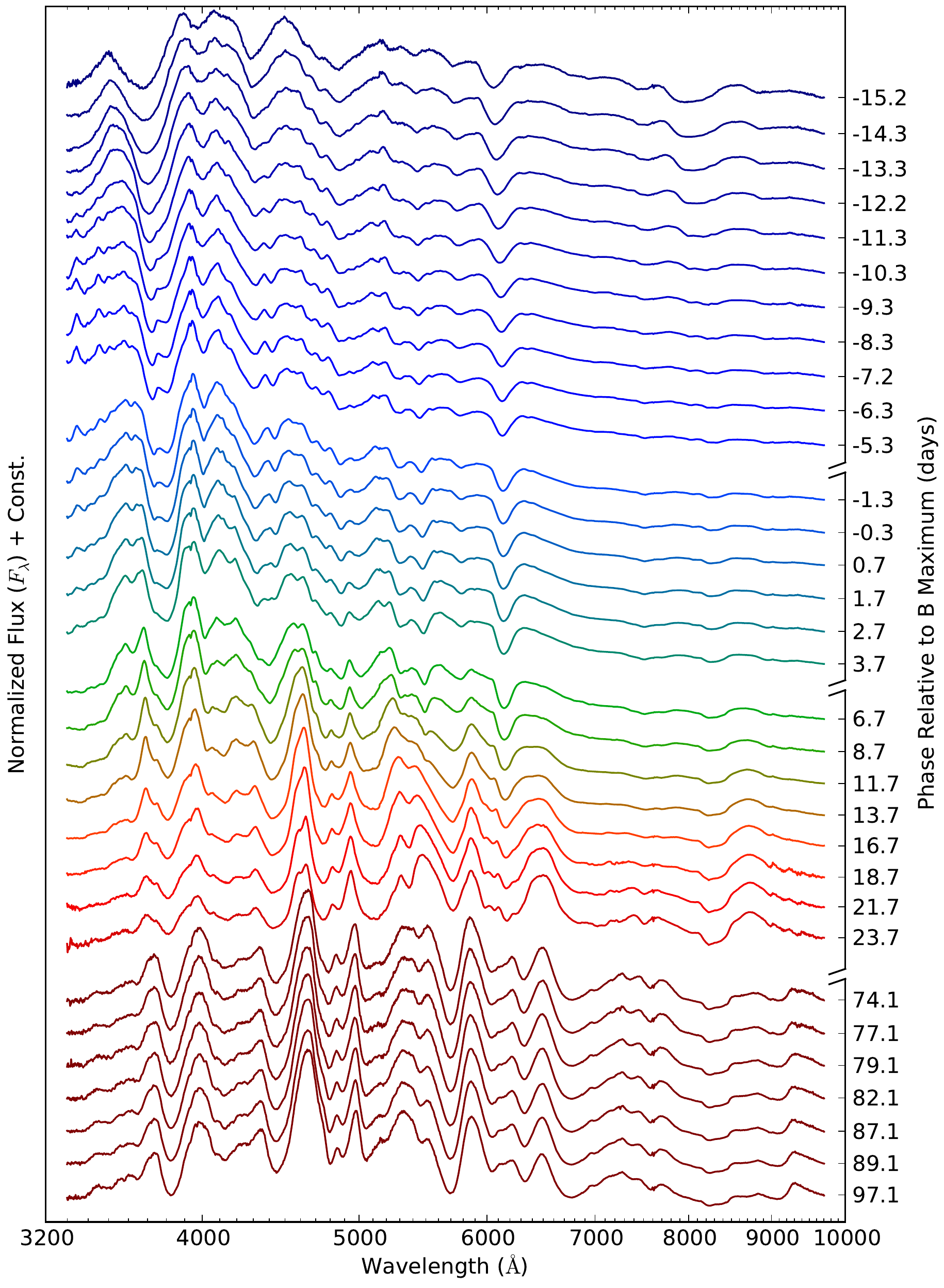}
    \caption{\snifs\ spectrophotometric time series of SN~2011fe from $-15$
      to 100~days relative to $B$-band maximum light.  Breaks in the axis on
      the right indicate gaps and changes to the observing cadence.  The
      first break corresponds to a four-day gap in daily cadence before
      maximum.  The second marks the change from daily to alternating
      two/three day cadence.  The final break is a 50 day hiatus imposed by
      lack of accessibility to the target from Mauna Kea.}
    \label{fig:timeseries}
\end{figure*}

The final spectrophotometric time series of SN~2011fe is depicted in
Fig.~\ref{fig:timeseries}.  The time series possesses a number of
outstanding qualities, beyond the extremely high S/N, despite the
calibration challenges.  Each spectrum spans the entire ground-based
optical wavelength window.  Excluding an inconvenient four day gap,
daily sampling is achieved on the rising side of the light curve.
Even though SN~2011fe observations were performed on non-standard
conditions when compared to the typical \snifs\ target, our observing
strategy and calibration procedure do an excellent job of removing
atmospheric artifacts.  These features, combined with the cosmological
utility of SN~2011fe as a ``normal'' \snia, should make these data a
highly useful resource.

Synthetic light curves were generated by integrating the product of
each spectrum with a set of non-overlapping top-hat filters with
perfect photon transmission in the following wavelength ranges:
3300--4102~\AA, 4102--5100~\AA, 5200--6289~\AA, 6289--7607~\AA, and
7607--9200~\AA, respectively \emph{UBVRI}$_\mathrm{SNf}$. This filter
set avoids the split between both \snifs\ spectrograph channels but is
contiguous otherwise. For convenience these measurements are given in
the Vega magnitude system by applying zero-points computed from the
latest Hubble Space Telescope (HST) spectral observation of
Vega~\citep{Bohlin07}, available from the CALSPEC
database.\footnote{\url{ftp://ftp.stsci.edu/cdbs/current_calspec/alpha_lyr_stis_005.fits}}
The synthesized photometry is listed in Table~\ref{tab:lc}, along with
the integrated flux over the whole \snifs\ optical window. For ease of
comparison with other SNe, we also present synthetic light curves
created using the latest photonic responses and zero-points given
by~\citet[BM12]{Bessell12}, which are denoted as
\emph{BVRI}$_\mathrm{BM12}$ in Table~\ref{tab:lc}. After interpolation
of those filter responses into 10~\AA\ steps using cubic splines, our
integrator agrees at the 0.001 mag level with the colors measured by
\citet{Bessell12} using the aforementioned Vega spectrum and the
zero-points derived by these authors.

The achromatic absolute flux calibration accuracy for each night as
estimated by the calibration pipeline is shown in the $\sigma$ column
of Table~\ref{tab:lc}. This value takes into account the errors of the
spectral extraction using the chromatic spatial PSF, estimated from
the mean scatter of flux calibrated standard star observations during
all the photometric nights of the \snf\ data set, as well as the
errors of the instrumental flux solution, the mean atmospheric
extinction and the photometric ratio computations. The main
contributions are the achromatic extraction uncertainty of 0.03 mag
and 0.02 mag empirically found for bright and faint standard stars
\citep{Buton09}, and a 0.03 mag additional uncertainty on
non-photometric nights due to the photometric ratios usage
\citep{Pereira08}. The chromatic spatial PSF flux extraction
efficiency was evaluated on residual spectra, extracted from the
PSF-subtracted datacubes by performing aperture photometry centered at
the position of the SN and within a radius of $3\sigma$ seeing. The
median missing flux for all exposures, on the 5 broad-band filters of
the \snf\ filter set, was found to be below the half-percent level for
all bands except for $U_\mathrm{SNf}$, where it is $\sim0.7$\%. A
$\la1$\% residual color effect is also discernible, when comparing
$B_\mathrm{SNf}$ with the other bands at high airmasses ($>2.7$), and
is due to the fact that at such atypically large airmass, the
atmospheric differential refraction (ADR) is so large that too much
light is projected out of the \snifs\ FOV. The contribution to the
total error from both residual chromatic effects is much smaller than
that from the absolute flux calibration.

\subsection{SALT2 fit}\label{sec:salt2-fit}

\begin{figure*}
  \centering
  \includegraphics[width=0.9\textwidth,clip=true]{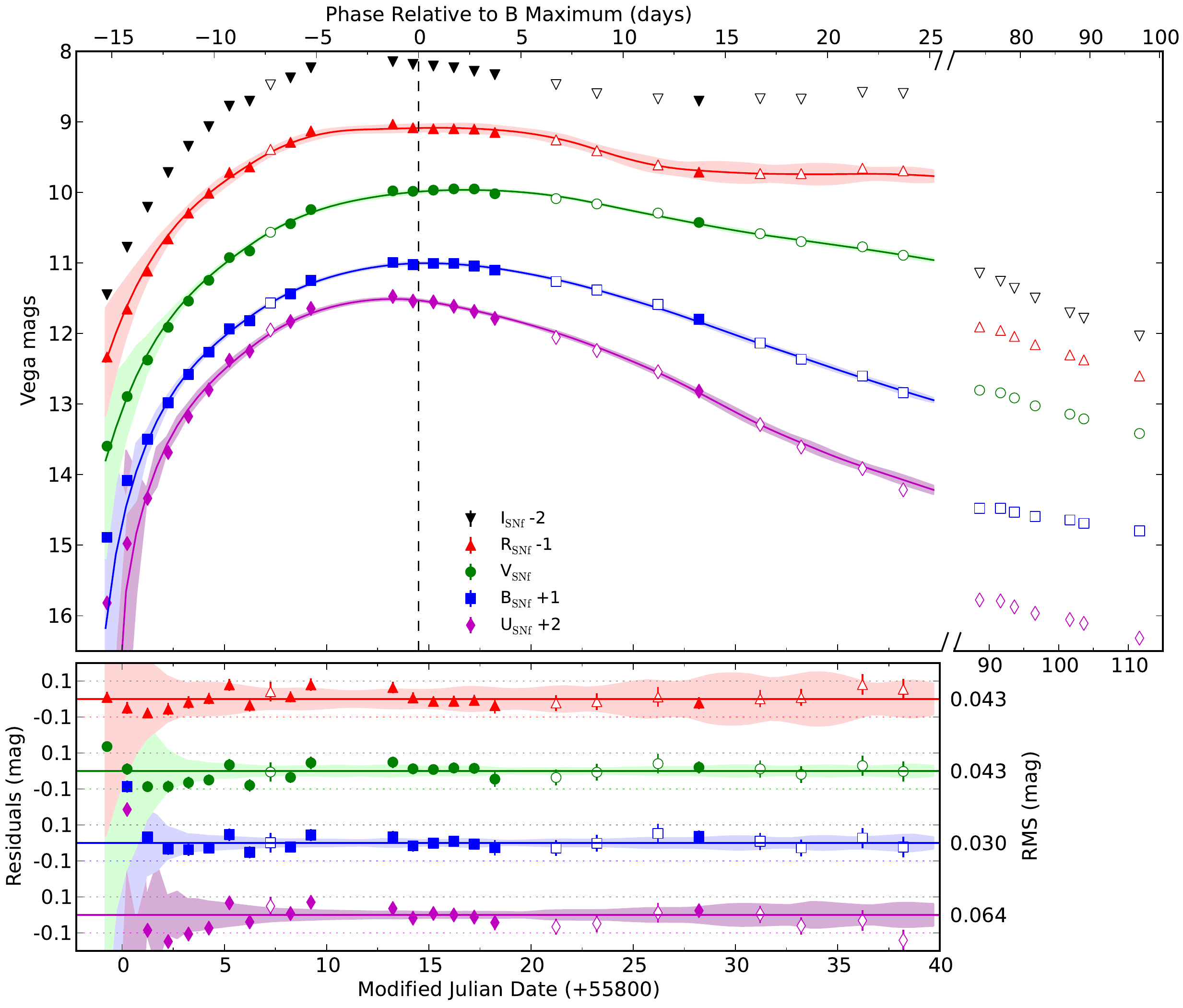}
  \caption{SN~2011fe synthesized light curves using the
    \emph{UBVRI}$_\mathrm{SNf}$ filter set. Filled and open symbols stand
    for photometric and non-photometric nights respectively. The
    results of a SALT2 simultaneous fit of \emph{UBVR}$_\mathrm{SNf}$ in
    the phase range $-\,16 < t < +\,25$~d are shown as solid lines,
    along with the corresponding residuals (SALT2 - \snifs) on the
    lower panel. The shaded areas represent the SALT2 model error. The
    residuals for the first points of U$_\mathrm{SNf}$ and B$_\mathrm{SNf}$
    fall outside the panel, and the RMS on the residuals for each band
    ignores the first 2 points. The break in the time axis corresponds
    to the $\sim50$ day gap in follow-up during which SN~2011fe was
    not visible during the night from Hawaii. Note the change of scale
    of the extended time axis covering the late observations.}
  \label{fig:lc}
\end{figure*}

The \emph{UBVR}$_\mathrm{SNf}$ light curves were simultaneously fitted using
SALT2.2~\citep[SALT2 henceforth]{Guy10}, for estimation of the date
of maximum light and assigning phases to the observations (relative to
maximum light).  The publicly available SALT2 spectral
model\footnote{\url{http://supernovae.in2p3.fr/~guy/salt/download_templates.html}}
seems to attain the (synthesized) $B$ peak slightly before phase 0,
which is a nuisance when we use it for a precise determination
of the date of maximum light of our light curve. The shift needed to
be applied to the fitted date of maximum, in order for it to match
with phase 0 in $B$, was determined empirically to be $\Delta {\rm
  DayMax}_{B} \approx -0.7$~d. The SALT2 code used here was modified
to take this shift into account.

For a coherent treatment of the \snf\ filter set by SALT2, which uses
a BD+17\;4708 based magnitude system, the following zero-points are
added to the
\verb\BD17-snls3.dat\ magnitude system configuration file: 9.787,
9.791, 9.353, 9.011, and 8.768; which represent BD+17\;4708's
synthetic magnitudes with relation to the Vega spectrum, in
\emph{UBVRI}$_\mathrm{SNf}$ respectively. Equivalently, if we intend to
use SALT2 to fit light curves synthesized using the \emph{UBVRI}$_{\rm
  BM12}$ photonic responses, the zero-points to be added to the
configuration file are 9.709, 9.902, 9.469, 9.163, and 8.843. It
should be stressed that every SALT2 fit result presented here is
in the ``standard'' SALT2 \emph{UBVRI} magnitude system, allowing
direct comparison with the literature. This magnitude system is based
on the \citet{Bessell90} filter transmissions, shifted to match
\citet{Landolt92} observations. If we use instead the magnitude system
based on the BM12 transmissions, the effect on the fitted parameters
is very small, and completely within the quoted uncertainties.

The fit was performed using all available observations within four
weeks of maximum light, including the very first one, which falls
just outside the default SALT2 phase range of $-\,15 < t < +\,45$~d.
Early observations are deweighted by the SALT2 error model.  Simply
omitting the first four observations (phases $<-\,12$~d) results in a
shift in the date of maximum of $-0.018$~d, well inside the quoted
error.

The light curves, fits and corresponding residuals can be seen in
Fig.~\ref{fig:lc} (for a comparison with light curve results already
published on the literature see Sect.~\ref{sec:light-curve-comp}).
The fitted $B$ maximum of $9.94 \pm 0.01 $ mag was reached on 2011
September 10.5.  The best-fit SALT2 parameters are $x_1 = -0.206 \pm
0.071, c = -0.066 \pm 0.021$ (see Table~\ref{tab:summary} for a full
summary of extracted photometric and spectroscopic parameters).
Excluding \emph{U}SNf-band data from the fit, we obtain a negligible
shift for the date and magnitude at maximum, and $x_1 = -0.149 \pm
0.096, c = -0.061 \pm 0.027$.  The light curve shape parameters are
typical of a ``normal'' (if slightly blue) \snia: the median values
for the $x_{1}$ and $c$ distributions of the nearby ($z\leq0.1$)
\sneia\ used by \citet{Conley11} are respectively $-0.249$ and
$-0.026$ (J.~Guy, private communication). The $V$-band absolute magnitude \citep[assuming
$\mu=29.04\pm0.19$,][]{Shappee11} at $B$ peak, $M_{\rm max_{B}}V =
-19.05\pm0.19$~mag matches perfectly the average found by
\citet[][$-19.06\pm0.05$~mag]{Riess09}.

The RMS of the residuals of the four fitted filters, ignoring the
first two exposures due to the large discrepancies with the SALT2
model (\cf\ Sect.~\ref{sec:template-comp}), are respectively 0.06,
0.03, 0.04 and 0.04 mag. The points that deviate the most from the fit
are those for phases $t<-10$ d, showing the inadequacy of the SALT2
model for such early phases. Nevertheless, the exposure of night MJD
55805 ($t~\sim-\,9$~d, \cf\ Fig.~\ref{fig:lc}) seems to present a
systematic error in its flux calibration of $\sim0.1$~mag, since it is
brighter than we would expect based on observations on neighboring
nights.

\begin{sidewaystable*}
  \caption{Synthetic light curves of SN~2011fe}
  \label{tab:lc}
  \centering
  {\footnotesize
    \begin{tabular}{crrrrrrrrrccc}
      MJD\tablefootmark{a} &
      $U_\mathrm{SNf}$ & $B_\mathrm{SNf}$ & $V_\mathrm{SNf}$ & $R_\mathrm{SNf}$ & $I_\mathrm{SNf}$ &
      $B_\mathrm{BM12}$ & $V_\mathrm{BM12}$ & $R_\mathrm{BM12}$ & $I_\mathrm{BM12}$ & $\sigma$\tablefootmark{b} &
      F(\snifs)\tablefootmark{c} & L(bolometric)\tablefootmark{d}\\
      & (mag) & (mag) & (mag) & (mag) & (mag) & (mag) & (mag) & (mag) &
      (mag) & (mag) & ($10^{-10}$ erg s$^{-1}$ cm$^{-2}$) & ($10^{42}$ erg s$^{-1}$)\\
      \hline\hline\\[-2ex]
      55799.3 & 13.819 & 13.888 & 13.596 & 13.333 & 13.451 & 13.871 & 13.611 & 13.514 & 13.565 & 0.030 & $\phn0.61 \pm 0.02$ & $\phn0.36 \pm 0.01$\\
      55800.2 & 12.977 & 13.085 & 12.893 & 12.654 & 12.777 & 13.054 & 12.890 & 12.821 & 12.876 & 0.033 & $\phn1.22 \pm 0.06$ & $\phn0.71 \pm 0.03$\\
      55801.2 & 12.336 & 12.497 & 12.376 & 12.115 & 12.210 & 12.455 & 12.377 & 12.278 & 12.297 & 0.030 & $\phn2.06 \pm 0.08$ & $\phn1.19 \pm 0.04$\\
      55802.3 & 11.686 & 11.983 & 11.912 & 11.659 & 11.719 & 11.924 & 11.913 & 11.804 & 11.804 & 0.034 & $\phn3.34 \pm 0.16$ & $\phn1.91 \pm 0.08$\\
      55803.2 & 11.175 & 11.579 & 11.540 & 11.292 & 11.347 & 11.510 & 11.549 & 11.421 & 11.420 & 0.034 & $\phn4.89 \pm 0.24$ & $\phn2.84 \pm 0.12$\\
      55804.2 & 10.798 & 11.263 & 11.246 & 11.011 & 11.068 & 11.190 & 11.258 & 11.130 & 11.129 & 0.031 & $\phn6.57 \pm 0.19$ & $\phn3.91 \pm 0.09$\\
      55805.2 & 10.379 & 10.935 & 10.924 & 10.716 & 10.778 & 10.857 & 10.943 & 10.816 & 10.832 & 0.034 & $\phn9.02 \pm 0.28$ & $\phn5.42 \pm 0.14$\\
      55806.2 & 10.249 & 10.817 & 10.830 & 10.639 & 10.707 & 10.739 & 10.847 & 10.732 & 10.752 & 0.034 & $\phn9.96 \pm 0.44$ & $\phn6.25 \pm 0.22$\\
      55807.3 & 9.950 & 10.568 & 10.565 & 10.390 & 10.476 & 10.488 & 10.589 & 10.472 & 10.518 & 0.054 & $12.72 \pm 0.63$ & $\phn7.86 \pm 0.31$\\
      55808.2 & 9.830 & 10.439 & 10.444 & 10.289 & 10.377 & 10.358 & 10.466 & 10.362 & 10.417 & 0.030 & $14.19 \pm 0.55$ & $\phn8.87 \pm 0.28$\\
      55809.2 & 9.644 & 10.248 & 10.243 & 10.128 & 10.234 & 10.167 & 10.266 & 10.181 & 10.274 & 0.034 & $16.80 \pm 0.53$ & $10.27 \pm 0.27$\\
      55813.2 & 9.475 & 9.993 & 9.978 & 10.033 & 10.147 & 9.913 & 10.001 & 9.992 & 10.214 & 0.032 & $20.09 \pm 0.82$ & $11.77 \pm 0.41$\\
      55814.2 & 9.540 & 10.023 & 9.984 & 10.084 & 10.186 & 9.946 & 10.006 & 10.019 & 10.269 & 0.031 & $19.38 \pm 0.79$ & $11.35 \pm 0.39$\\
      55815.2 & 9.552 & 10.006 & 9.968 & 10.098 & 10.209 & 9.936 & 9.990 & 10.011 & 10.308 & 0.030 & $19.39 \pm 0.77$ & $11.23 \pm 0.39$\\
      55816.2 & 9.614 & 10.007 & 9.950 & 10.098 & 10.234 & 9.943 & 9.974 & 10.001 & 10.340 & 0.031 & $19.07 \pm 0.75$ & $10.93 \pm 0.37$\\
      55817.2 & 9.687 & 10.045 & 9.951 & 10.103 & 10.283 & 9.983 & 9.981 & 10.002 & 10.388 & 0.031 & $18.40 \pm 0.72$ & $10.41 \pm 0.36$\\
      55818.2 & 9.786 & 10.101 & 10.019 & 10.150 & 10.332 & 10.049 & 10.043 & 10.065 & 10.439 & 0.042 & $17.26 \pm 0.93$ & $\phn9.64 \pm 0.46$\\
      55821.2 & 10.055 & 10.264 & 10.088 & 10.255 & 10.472 & 10.232 & 10.116 & 10.171 & 10.585 & 0.044 & $14.87 \pm 0.61$ & $\phn8.13 \pm 0.30$\\
      55823.2 & 10.241 & 10.384 & 10.163 & 10.408 & 10.601 & 10.370 & 10.189 & 10.299 & 10.739 & 0.047 & $13.18 \pm 0.57$ & $\phn7.12 \pm 0.28$\\
      55826.2 & 10.542 & 10.587 & 10.293 & 10.610 & 10.675 & 10.607 & 10.318 & 10.491 & 10.879 & 0.054 & $10.99 \pm 0.55$ & $\phn5.84 \pm 0.27$\\
      55828.2 & 10.817 & 10.800 & 10.426 & 10.712 & 10.707 & 10.845 & 10.462 & 10.622 & 10.933 & 0.034 & $\phn9.35 \pm 0.41$ & $\phn4.99 \pm 0.20$\\
      55831.2 & 11.290 & 11.135 & 10.585 & 10.732 & 10.673 & 11.231 & 10.657 & 10.717 & 10.897 & 0.048 & $\phn7.59 \pm 0.33$ & $\phn4.11 \pm 0.16$\\
      55833.2 & 11.609 & 11.365 & 10.697 & 10.733 & 10.679 & 11.490 & 10.792 & 10.768 & 10.873 & 0.047 & $\phn6.66 \pm 0.29$ & $\phn3.67 \pm 0.14$\\
      55836.2 & 11.915 & 11.603 & 10.771 & 10.657 & 10.583 & 11.758 & 10.900 & 10.749 & 10.758 & 0.057 & $\phn6.16 \pm 0.32$ & $\phn3.43 \pm 0.16$\\
      55838.2 & 12.216 & 11.838 & 10.891 & 10.692 & 10.598 & 12.007 & 11.038 & 10.818 & 10.757 & 0.057 & $\phn5.50 \pm 0.29$ & $\phn3.14 \pm 0.14$\\
      55888.6 & 13.776 & 13.479 & 12.804 & 12.906 & 13.145 & 13.608 & 12.953 & 12.894 & 13.044 & 0.044 & $\phn0.91 \pm 0.04$ & \nodata\\
      55891.6 & 13.790 & 13.479 & 12.841 & 12.956 & 13.260 & 13.606 & 12.986 & 12.939 & 13.131 & 0.045 & $\phn0.88 \pm 0.04$ & \nodata\\
      55893.6 & 13.875 & 13.531 & 12.913 & 13.041 & 13.360 & 13.670 & 13.052 & 13.021 & 13.229 & 0.044 & $\phn0.83 \pm 0.03$ & \nodata\\
      55896.6 & 13.966 & 13.595 & 13.025 & 13.158 & 13.498 & 13.736 & 13.157 & 13.138 & 13.357 & 0.043 & $\phn0.76 \pm 0.03$ & \nodata\\
      55901.6 & 14.054 & 13.643 & 13.143 & 13.303 & 13.708 & 13.789 & 13.257 & 13.275 & 13.546 & 0.042 & $\phn0.69 \pm 0.03$ & \nodata\\
      55903.6 & 14.108 & 13.691 & 13.210 & 13.374 & 13.784 & 13.839 & 13.320 & 13.347 & 13.618 & 0.041 & $\phn0.65 \pm 0.02$ & \nodata\\
      55911.6 & 14.317 & 13.798 & 13.417 & 13.601 & 14.036 & 13.961 & 13.500 & 13.576 & 13.869 & 0.039 & $\phn0.55 \pm 0.02$ & \nodata\\
      \hline
    \end{tabular}}
  \tablefoot{Synthetic photometry in Vega magnitudes. The
    \emph{UBVRI}$_\mathrm{SNf}$ magnitudes are synthesized using
    non-overlapping top-hat filters with perfect photon transmission in
    the following wavelength ranges: 3300--4102~\AA, 4102--5100~\AA,
    5200--6289~\AA, 6289--7607~\AA, and 7607--9200~\AA. The
    \emph{BVRI}$_\mathrm{BM12}$ magnitudes are synthesized using photonic
    responses and zero points from ~\citet{Bessell12}.\\
    \tablefoottext{a}{JD - 2400000.5}\\
    \tablefoottext{b}{Achromatic estimated $1\sigma$ flux calibration
      uncertainties, which largely dominate over photon shot noise.}\\
    \tablefoottext{c}{Total flux integrated over the whole \snifs\
      wavelength range.}\\
    \tablefoottext{d}{The near-infrared measurements by \citet{Matheson12} do not
      cover dates beyond 45 days after maximum light.}}
\end{sidewaystable*}


\section{Analysis}\label{sec:analysis}

\subsection{Light curve comparison}\label{sec:light-curve-comp}

The optical light curves of SN~2011fe published by different authors
allow us to assess the reciprocal precision of multiple independent
observations of this SN, and ultimately the benefits of using \sneia\
spectrophotometry for synthetic light curve creation. We use for these
comparisons the first set of \snf\ observations, up to MJD 55840. For
every \emph{comparison} light curve provided by other followup
campaigns, the approach used to compute the residuals with respect to
the \emph{reference} \snifs\ is the same: a light curve is synthesized
from our spectrophotometric time series in the same passband as the
comparison light curve, and then fit using gaussian processes
\citep[as implemented by \texttt{scikit-learn};][]{sklearn} employing
a squared euclidean correlation model weighted by measurement
errors. The fitted model is then used to evaluate the synthesized
\snifs\ light curve at the same observation dates of the comparison
light curves. The magnitude residuals are always \emph{reference}
(\snifs) minus \emph{comparison}. Positive or negative residuals thus
mean that the comparison observation is respectively brighter or
fainter than the reference.

\begin{figure*}
    \centering
    \includegraphics[width=0.9\textwidth,clip=true]{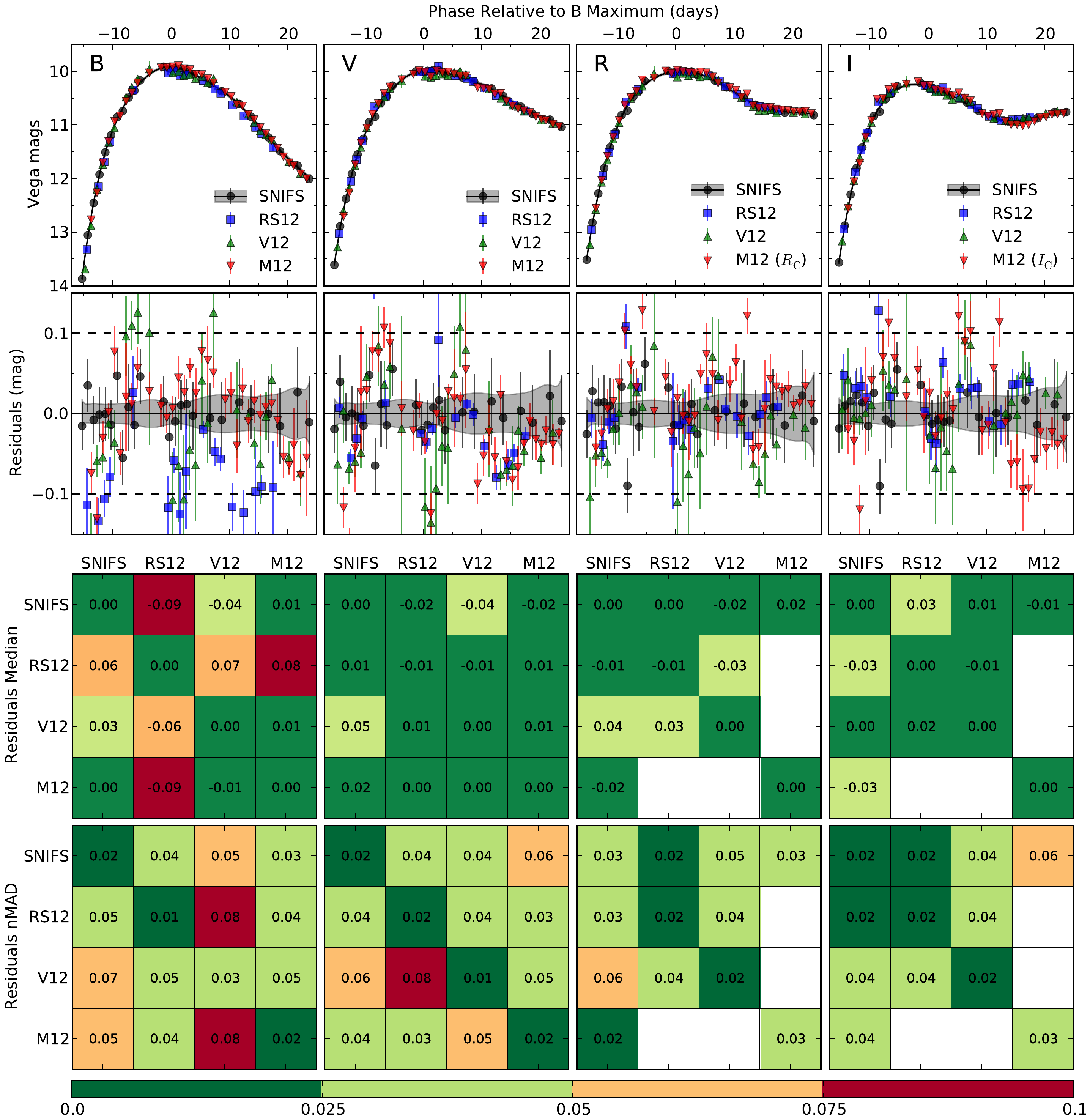}
    \caption{First panel: comparison of synthesized \emph{BVRI}
      \snifs\ light curves of SN~2011fe, using BM12 passbands and zero
      points, with published photometry from
      \citet[][RS12]{Richmond12}, \citet[][V12]{Vinko12} and
      \citet[][M13]{Munari13}. Second panel: residuals between all
      light curves and the \snifs\ interpolated light curve, whose
      uncertainty is represented as the gray shaded region. Third \&
      forth panel: median and normalized median absolute deviation of
      the residuals for each combination of \emph{reference} (row) and
      \emph{comparison} (column) light curves. The statistics of the
      residuals shown in the second panel, with \snifs\ as reference,
      are on the topmost row of each matrix. The
      $R_\mathrm{C}I_\mathrm{C}$ filters used by M13 are not directly
      comparable to \emph{RI} from RS12 and V12.}
    \label{fig:lcresiduals}
\end{figure*}

We start with a brief comparison with the space-based
\emph{Swift}/UVOT observations by \citet[][B12]{Brown12}, before
performing an extensive comparison with the three ground-based
photometric followup compaigns who published observations of
SN~2011fe: \citet[][RS12]{Richmond12}, \citet[][V12]{Vinko12} and
\citet[][M13]{Munari13}. This trio of light curves and residuals with
relation to synthesized \snifs\ photometry using BM12 passbands and
zero points, are shown in the two upper panels of
Fig.~\ref{fig:lcresiduals}. While a quick look at the superposed light
curves may give the impression of a good agreement between all
experiments, a careful study of the residuals shows that this is not
always the case. The median and normalized median absolute deviation
(nMAD) of the residuals between all the experiments are summarized in
the matrices shown in the lower two panels of
Fig.~\ref{fig:lcresiduals}. These matrices relate the statistics of
the residuals (observational bias and scatter) between a specific data
set and all the other ones, where each row uses a common
\emph{reference} and each column represents a different
\emph{comparison} data set. The bias and scatter matrices are not
perfectly (anti-)symmetric since that the interpolation of the light
curve depends on the time sampling and measurement errors of the
reference data set, allied to the fact that the nMAD is less dependent
on outliers. The statistics of the residuals with relation to \snifs,
plotted on the second panel of Fig.~\ref{fig:lcresiduals}, can be read
from the topmost row of each matrix.

\subsubsection{\citet{Brown12}}
\label{sec:brown12}

B12 present measurements in several UVOT filters, of which only $b$
and $v$ \citep{Poole08} fully overlap with the \snifs\ wavelength
range. Unfortunately both filters saturated before maximum light, and
as such our comparison is limited to phases $t < -\,8$~d and $+\,18 <
t < +\,25$~d. We use the most recent UVOT effective area curves and
zero points \citep{Breeveld11} and integrate our spectra in
$F_{\nu}$. This approach was validated using a Vega spectrum, from
which we found exactly the same zero points as reported by
\citet{Breeveld11}. We found a clear bias on the residuals, whose
median values are $-0.14$ and $-0.20$~mag respectively for $b$ (with
comparison dates only at $t<-\,8$~d) and $v$, with no evident
correlation with magnitude. This bias also seems to be present for the
bluer filters (\cf\ Sect.~\ref{sec:bolometric-lc}). The cause for this
effect is not clear. The photometric calibration of UVOT is a
notoriously difficult endeavor \citep{Poole08}, and the fact that the
instrument nears the saturation limit for SN~2011fe could point to
problems in the coincidence loss or aperture
corrections. $S$-corrections \citep{Suntzeff00} may also play a role,
since \citet{Brown12} use the \citet{Hsiao07} template to convert the
observed count rates into flux, while these template spectra do not
exactly match the observed ones at early phases (\cf\
Sect.~\ref{sec:template-comp}).

\subsubsection{\citet{Richmond12}}

RS12 present \emph{BVRI} observations coming primarily from the
Rochester Institute of Technology (RIT) observatory. These
observations are the most extensive and consistently calibrated from
the full AAVSO followup sample used by \citet{Tammann11}. The biases
for \emph{VRI} are small and within the quoted errors of the synthetic
photometry presented in this work, while the scatters are smaller or
similar to the ones found for the SALT2 fit. The $B$-band however
shows a large bias with relation to \snifs\ observations
($\sim0.1$~mag), due to an $S$-correction problem with the calibration
of RS12. That is evidenced by the unusually large color coefficient
derived by the authors for this filter ($0.24 \pm 0.04$), and by the
fact that when using the RS12 $B$-band light curve as \emph{reference}
(second row of the first matrix), we find large biases with relation
to every other data set. The CCD used in the RIT observatory has a
very pronounced quantum efficiency (QE) drop in the $B$-band
wavelength range, which explains this effect. Using CALSPEC spectra
for stars with similar colors to the ones from the PG1633+099 standard
field used by RS12 for their calibration, the BM12 passbands convolved
by the KAF--1600 CCD QE curve, and an atmospheric extinction model
\citep[\texttt{pyExtinction};][]{Buton13} based on typical atmospheric
values for the RIT site, we obtain a $b-v$ color coefficient of $0.31
\pm 0.10$, while that ignoring the QE effect gives $-0.07 \pm
0.08$. The QE convolved passbands also improve the correlation of the
$B$-band synthetic photometry with the instrumental
$b_\mathrm{SN}-b_\mathrm{Vega}$, after inversion of the color
correction performed by RS12. Differential ADR between a blue \snia\
and red reference stars will also affect filter photometry at some
level, especially at the high airmasses of SN~2011fe observations. The
authors' derived values for the $B$-band date of maximum, peak
magnitude and decline rate are then naturally different from ours,
being greater by approximately 1 day, 0.05 mag and
0.1\footnote{\citet{Vinko12} attribute this difference to a probable
  misprint, and find $\Delta m_{15} (B) = 1.12\pm0.05$ using RS12
  data. Fitting the same data with SALT2 we find $1.17\pm0.04$.}
respectively. We point out that the same values derived by
\citet{Tammann11} using the full AAVSO sample are perfecly compatible
with those found in the present work.

\subsubsection{\citet{Vinko12}}

V12 published \emph{BVRI} photometry obtained at the Konkoly
Observatory. The \emph{BI} bands are more compatible with \snifs\
observations than their RS12 counterparts, but this is not the case
for \emph{VR}. The CCD that was used has a less steep QE curve in the
bluer wavelengths, and is thus less prone to $S$-correction problems
as shown by the smaller bias for the $B$-band, still $\sim0.04$~mag
fainter than the \snifs\ photometry. This bias does not seem to be
present if we use M13 (\cf\ \ref{sec:munari13}) instead as the
reference light curve, which leads us to think that the bias is
affected by the quality of the interpolated light curve, and its
dependence on individual point measurement errors. Globally, V12 data
shows some tension when used as reference, especially with \snifs\ for
\emph{BVR}. The scatter of the residuals with relation to \snifs\ is
of the order of $0.05$~mag for all bands, which is higher than RS12,
especially for the redder filters. The $B$-band date of maximum and
decline rate obtained by V12 using MLCS2K2 are in accord with our own.

\subsubsection{\citet{Munari13}}
\label{sec:munari13}

Finally, M13 published $BVR_\mathrm{C}I_\mathrm{C}$ photometry obtained
using several telescopes of the Asiago Novae and Symbiotic stars
collaboration \citep[ANS;][]{Munari12}. They acknowledge the
importance of $S$-corrections when doing \sneia\ photometry with
multiple instrumental setups: in addition to accurate photometric
calibration sequences \citep{Henden12}, M13 implement a light curve
merging method that uses phase-dependent zero points for each
telescope, found via a global $\chi^{2}$ minimization per band. For
the comparisons with the $R_\mathrm{C}I_\mathrm{C}$ photometry, which is
different from that synthesized using BM12's \emph{RI}, we used the
\citet{Landolt92} passbands adjusted to photon count transmission, in
accord with the standard star catalogs used for calibration by
\citet{Henden12}. M13 do not publish individual measurement errors,
but present instead the calibration error for each band. This error is
smaller than the merged light curve individual night scatter, when
multiple observations are available. We thus average all same night
observations, and use as a conservative error the error on the mean of
those same observations added quadratically to the corresponding band
calibration error. The error for single observation nights is set to
the average of the errors of all multiple observation nights. The
$B$-band bias and scatter with relation to \snifs\ are the lowest of
the three photometric data sets considered, attesting to the quality
of their calibration procedure: M13 is a very good reference for all
the $B$ data sets not affected by $S$-correction problems. The biases
in the other bands are comparable and do not exceed
$\sim0.02$~mag. The observational scatters are slightly larger for
\emph{VI} probably due to outliers in one of the 3 merged telescope
data sets. The $B$-band date of maximum and peak magnitude found by
M13, for which they do not quote any error estimate, are very close to
our own (within 0.2~d and 0.02~mag), while the decline rate is
perfectly compatible within our estimated error.

\vspace{.5em}

The synthetic light curves derived from the spectrophotometric time
series presented in this paper were compared with those in the
literature. Most of the data sets have the smallest residual biases in
$V$, and show an intrinsic scatter of 0.02--0.03~mag for all bands.
The typical scatter between data sets comes mostly from calibration
issues, and is of the order of 0.03--0.06~mag, similar to the SALT2
fit residuals and with no difference when the comparison is made
between purely photometric data sets or with \snifs. This highlights a
major benefit of the \snf\ data, which is at the same time impervious
to $S$-correction problems due to its spectophotometric nature, and of
photometric quality comparable to traditional photometric followup
data sets, even under the extreme observational conditions pertaining
to SN~2011fe.

\subsection{Interstellar absorption in M101}
\label{sec:dust}

Estimates of reddening by the host galaxy may be obtained from either
photometric or spectral data.  We consider our photometric data first.
\citet{Folatelli10} reapplied the procedure used by \citet{Lira96} and
\citet{Phillips99} and derived an intrinsic color law for the ``tail''
of \snia\ light curves, and for maximum light using pseudocolors (the
difference between the magnitudes of two bands at each band's date of
maximum).  Their color law uses measurements on the CSP photometric
system, which we synthesize from our spectra using the filters and
zero-points given by \citet{Stritzinger11}. The light curves obtained
are interpolated using SALT2 and two independent fits on
\emph{BV}$_\mathrm{CSP}$. We measure $\Delta m_{15} (B_{\rm
  CSP})=1.056\pm0.035$ and obtain E$(B-V)_\mathrm{max}=0.002 \pm 0.062$
mag, by applying Equation~3 of \citet{Folatelli10}. We added 0.060 mag
in quadrature to the error, to account for the dispersion of the fit
by \citet{Folatelli10}. Applying their Equation~2 to the three spectra
whose phase is $+\,30 < t < +\,80$~d with respect to $V_\mathrm{CSP}$
maximum, we find a weighted mean and standard deviation of
E$(B-V)_\mathrm{tail}=0.038 \pm 0.045$ mag. Each individual measurement
error was increased in quadrature by the dispersion of the
corresponding fit, 0.077 mag. Both values are compatible and agree
with the trend found by \citet{Folatelli10} between the difference of
both measurements and E$(B-V)_\mathrm{max}$ (upper-left panel of their
Fig. 12).  Using a weighted mean of these estimates, and assuming they
are independent, we find $\left<\mathrm{E}(B-V)_{\rm
    host}\right>=0.026\pm0.036$ mag for the reddening due to dust on
the line of sight to SN~2011fe in M101.  This is compatible with the
value ($0.03 \pm 0.06$ mag) found by \citet{Tammann11} using \snia\
intrinsic colors derived empirically by \citet{Reindl05}.

Spectroscopically, we measured simultaneously the equivalent width of
\ion{Na}{i}~D for the Milky Way (MW) and M101 using fixed gaussian
profiles with a doublet ratio of $2\!\!:\!\!1$. We obtain respectively
EW(\ion{Na}{i}~D)$_\mathrm{MW}=10^{+22}_{-46}$ m\AA\ and
EW(\ion{Na}{i}~D)$_\mathrm{host}=-8^{+18}_{-56}$ m\AA, with a 95\%
confidence limit at 162 m\AA.  The latter is consistent with the value
found by \citet{Nugent11} using HiRES ($45\pm9$ m\AA) and both results
are also compatible with high resolution measurements by
\citet{Patat11}, who find $38\pm5$ m\AA\ and $47\pm2$ m\AA\
respectively. Using the empirical relation proposed by
\citet{Poznanski12} to derive the dust extinction from the $\rm
D_{1}+D_{2}$ \ion{Na}{i}~D lines, we obtain E$(B-V)_{\rm
  host}=0.014\pm0.003$~mag, in accord with our photometrically derived
value. Here we find only a statistical uncertainty, as the systematic
errors are difficult to quantify.

In conclusion, we confirm there is little evidence for significant
extinction of SN~2011fe by dust in its host galaxy.  Therefore, no
reddening corrections are performed on the \snifs\ spectra, other than
for MW extinction.

\subsection{Bolometric light curve and $^{56}\mathrm{Ni}$ mass}
\label{sec:bolometric-lc}

To construct a bolometric light curve of SN~2011fe, we combine \snifs\
optical spectrophotometry with ultraviolet (UV) and near-infrared
(NIR) \snia\ spectral templates.

\begin{figure}
  \centering
  \includegraphics[width=0.49\textwidth,clip=true]{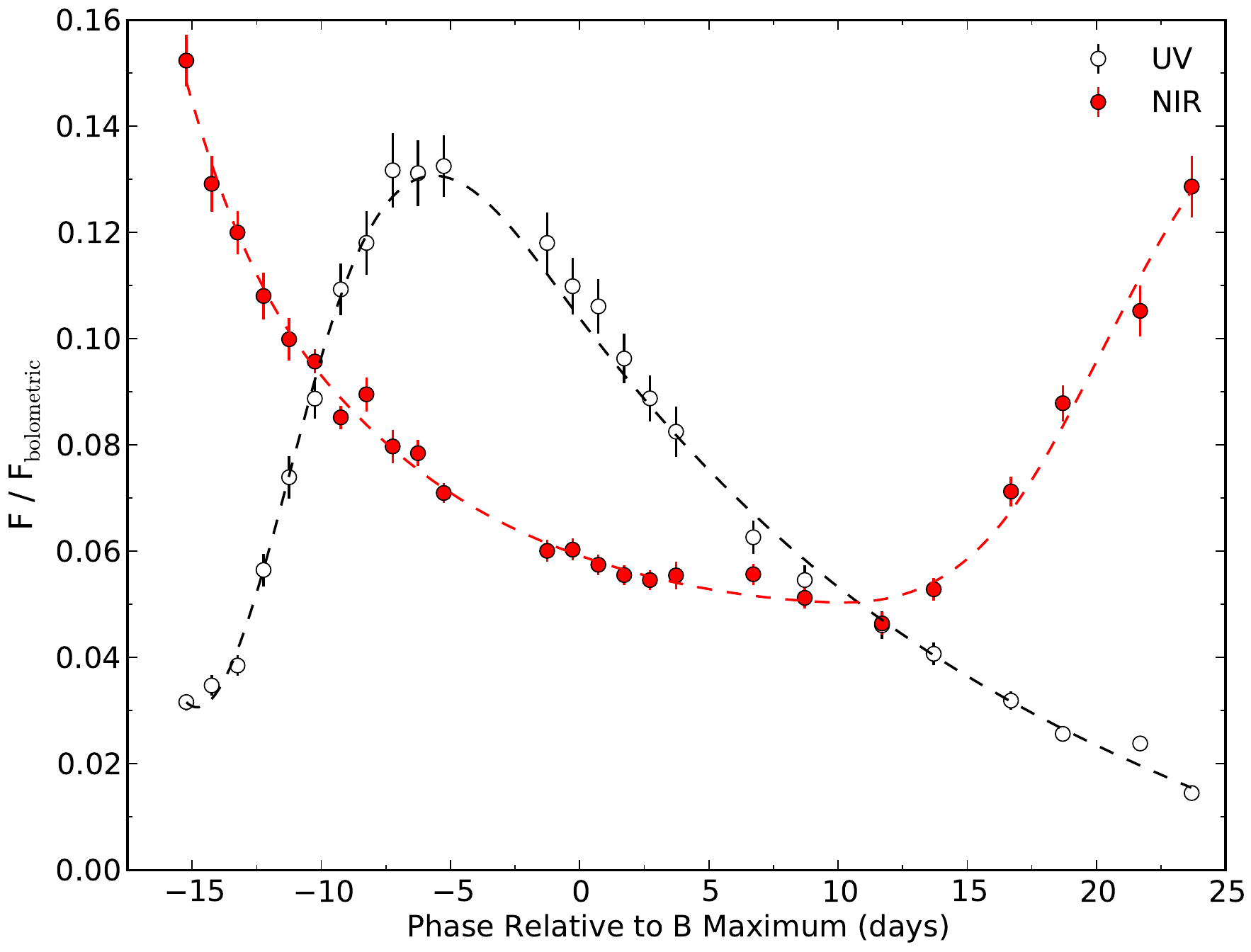}
  \caption{Phase evolution of the ratio of flux in the ultraviolet
    (1600--3400~\AA) and near-infrared (9700--24000~\AA) to total
    bolometric flux for SN~2011fe. The dashed lines represent cubic
    spline fits.\label{fig:uvnir}}
\end{figure}

The UV template (1600--3400~\AA) was constructed from 25 HST STIS
spectroscopic observations of SN~2011fe, encompassing phases $-\,15
\leq t < +\,24$~d. All observations are part of the proposal GO-12298
\citep{Ellis09} and are publicly available from the HST archive. All
spectra were taken using the $52\times0.2\arcsec$ long-slit aperture and one
of the UV (\emph{G230L} or \emph{G230LB}) or the \emph{G430L}
gratings, together covering $\sim1600$--5700~\AA. Same-night
observations were averaged, and spectra from different gratings merged
together by averaging over their common wavelength range. The ensemble
was cut between 1700--3400~\AA\ and set to decrease linearly to zero
flux at 1600~\AA. Reduction artifacts identified as big isolated
``spikes'' in the spectra were removed. A multi-phase SN~2011fe UV
template was then created, by linear extrapolation of the time
evolution of the flux per wavelength bin, for the original 25 phases
plus a null spectrum at $-\,20$~d, to represent the pre-explosion
phase. For each phase of SN~2011fe we intend to reproduce, the
template is (achromatically) flux normalized with respect to the
\snifs\ spectrum at their common wavelength range (3300--3400~\AA), in
order to account for miscalibrations of the HST data. The median of
the residuals of synthetic photometry performed on the final
template, with respect to the measurements reported by \citet{Brown12}
for \emph{Swift}/UVOT \emph{uvw2}, \emph{uvm2} and \emph{uvw1}, are
respectively: $-0.22$, $-0.21$ and $-0.20$~mag. These values are
similar to what was found in \ref{sec:brown12}, once again implying
that the UVOT SN~2011fe measurements suffer from large biases. These
residuals should be taken with caution however, since they are very
dependent on the blueward flux distribution of the HST spectra, where
the noise is high and our template is therefore less
accurate. Nevertheless, since the UV flux of a normal \sneia\
represents only a small fraction of the total flux (\cf\
Fig.~\ref{fig:uvnir}), the impact of using a crude UV template for the
bolometric flux will be modest. The variance of the template is
constructed so that we have a 0.03~mag error on the full integrated UV
flux.

For the NIR (9700--24000~\AA), the \citet{Hsiao07} \snia\ spectral
template was used as a starting point. After extrapolation to each
phase, the template is chromatically warped in order to match the
SN~2011fe $JHK$ observations of \citet{Matheson12} from WIYN/WHIRC. A
quadratic spline, constructed from the ratio of observed over
synthetic photometry for each of the effective wavelengths of the
three WHIRC filters, is used. The variance is constructed from a flat
error estimation and warped chromatically in a similar way, in order
to match the \citet{Matheson12} measurement errors. The median
residuals and scatter of the final NIR template with relation to
observations are at or below the half percent level.

The bolometric flux is obtained by integrating the full wavelength
range 1600--24000~\AA\ of the UV + \snifs\ + NIR spectra. The
evolution of the ratio of UV and NIR flux to total flux is shown in
Fig.~\ref{fig:uvnir}. The UV flux accounts for a few percent of the
total flux with a maximum contribution of about 13\% attained five
days before $B$-band maximum. It then decreases steadily to reach
about 2\% at $t>+\,20$~d. The NIR contribution starts at about 15\% at
very early phases and declines to $5$\% around 10 days after $B$
maximum, increasing once again to reach 15--20\% at the time of the
secondary maximum in the NIR ($t\sim30$~d), that our data did not
sample.

In order to study the time evolution of integrated filter photometry
relative to the bolometric flux, we simulated the ratio to total flux
of individual or combined optical passbands, using the BM12
throughputs. For the full phase range considered here, the individual
band with the least scatter on the ratio of bolometric flux is $I$,
even if the mean ratio to total flux is small ($7.3\% \pm 2\%$). The
mean flux ratio on $V$ is higher while keeping a relatively small
scatter ($15.7\% \pm 3.1\%$), similar to what was seen by
\citet{Wang09a} for SN~2005cf. If we sum fluxes of multiple bands, the
most promising combinations are those including bands from both
extremes of the optical window, namely \emph{UVI} ($36.7\% \pm 1.3\%$)
and \emph{URI} ($38.0\% \pm 1.5\%$). The scatter of the \emph{URI}
combination decreases below the percent level if one limits phases to
$t<\,15$~d. The best non-$U$ combination is \emph{BI}, with mean ratio
and scatter of $32.1\% \pm 1.9\%$. Assuming the relative normality of
SN~2011fe, these ratios can thus be used to compute estimates of total
bolometric flux from near maximum optical measurements of a typical
\snia.

\begin{figure}
  \centering
  \includegraphics[width=0.49\textwidth,clip=true]{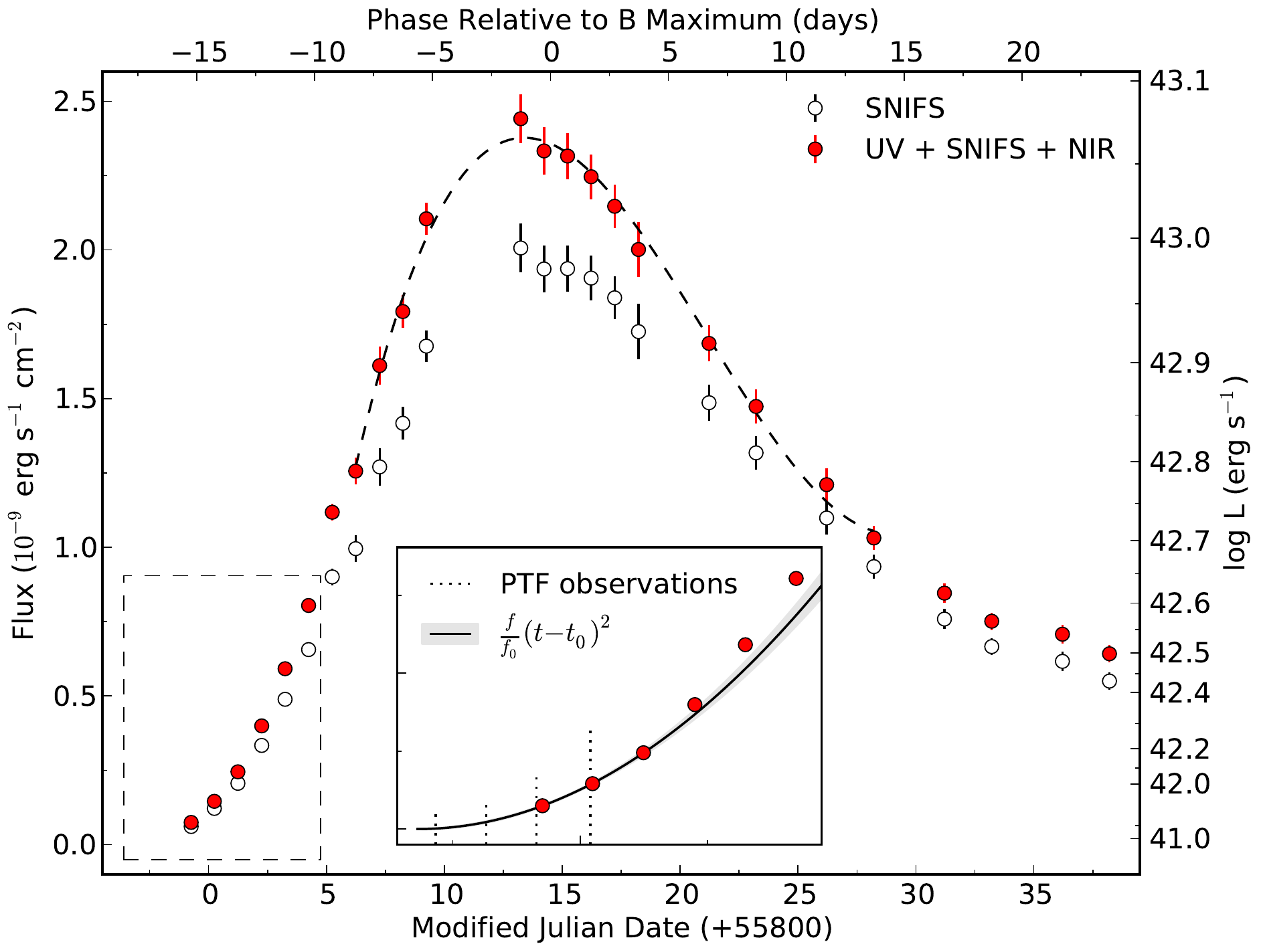}
  \caption{Bolometric light curve for SN~2011fe (filled red
    circles). The open circles represent the integrated flux over the
    full \snifs\ wavelength range and the dashed line is a cubic
    spline fit for determination of the date of maximum. The inset is
    a zoom of the area marked by a dashed box, where a quadratic rise
    is fit to the first 3 nights of \snifs\ observations. The vertical
    dotted lines mark the time of observations presented by
    \citet{Nugent11}.\label{fig:bolometric}}
\end{figure}

The distance-corrected \citep{Shappee11} bolometric luminosity is
presented in the last column of Table~\ref{tab:lc}, and the bolometric
flux is plotted in Fig.~\ref{fig:bolometric}, along with the
integrated flux over the full \snifs\ optical window. A $t^{2}$ power
law was fitted to our first three observations (inset of
Fig.~\ref{fig:bolometric}), under the assumption that the luminosity
dependence is dominated by the evolution of the photosphere surface
area \citep{Riess99,Goldhaber01,Gonzalez-Gaitan12,Piro12} until about
five days after the explosion. We find an explosion date (MJD)
$t_{0}=55796.81\pm0.13$, where the $\sim3$h precision is due to the
lack of earlier \snifs\ observations. The value is compatible with the
one found by \citet{Nugent11} ($55796.696\pm0.003$) using $g$-band
observations which start two days before the \snf\ ones. The goodness
of the fit decreases if we include the fourth night of \snifs\
observations, supporting the hypothesis that the $t^{2}$ model is only
valid until $\sim5$~d after the explosion. By letting the exponent of
the power law float we find a best-fit exponent of $2.21\pm0.51$ and
$t_{0}=55796.47\pm0.83$. This is a fully constrained fit with 3
parameters on 3 points, and earlier observations would be needed for
better precision. The former ($t^{2}$) analysis made using the $V_{\rm
  BM12}$ band yields $t_{0}=55796.68\pm0.12$, in closer accordance
with the PTF value, the one found by \citet{Brown12} when using the
$v$ band of UVOT ($55796.62\pm0.03$), and by \citet{Vinko12}
($55796.70\pm0.16$) using $R$-band observations. The derived data of
explosion thus seems to be dependent on the wavelengths that are
included in the bolometric flux.

By fitting a cubic spline to the bolometric light curve in the phase
range $-\,9 < t < +\,15$~d, we find the date of maximum bolometric
luminosity and derive a bolometric rise time $\tau_{r,\,bol}=t_{{\rm
    max},\,bol}-t_{0}=16.58 \pm 0.14$~d, where we use the explosion
date found with the bolometric flux and the $t^{2}$ power law, and
assume an error on the determination of the date of maximum of 0.06~d,
as with the SALT2 fits. Using the date of $B$-band maximum from
Sect.~\ref{sec:salt2-fit}, we obtain $\tau_{r,\,B}=17.70 \pm 0.14$~d,
1.1~d more than the bolometric rise time and compatible with the
\snia\ sample of \citet{Contardo00}, who found both rise-times to be
within 2 days. One should notice that those authors used
pseudo-bolometric light curves from integrated \emph{UBVRI}, most
closely matched by our \snifs\ integrated flux, from which we find a
date of maximum 0.5~d later than $\tau_{r,\,bol}$ and hence closer to
$\tau_{r,\,B}$. Employing similarly integrated \snifs\ data,
\citet{Scalzo12} found a difference of $\sim1$~d for a sample of
overluminous \sneia. Our stretch-corrected $\tau^{\prime}_{r,B}=18.27
\pm 0.14$~d is in agreement with the median value found by
\citet{Gonzalez-Gaitan12} for the SNLS sample when using SALT2 ($18.16
\pm 0.44$~d).

The maximum bolometric luminosity of SN~2011fe obtained from the
spline fit is L$_{\mathrm{max},bol} =
(1.17\pm0.04)\times10^{43}$~erg~s$^{-1}$. This can be used along with
the light curve rise-time to compute the $^{56}\mathrm{Ni}$ mass
synthesized in the explosion. Different authors used different
rise-time estimates, either from (pseudo) bolometric or integrated
band light curves. As already shown, these estimates will differ
slightly, thus impacting the computed $^{56}\mathrm{Ni}$ mass. We
chose to use our bolometric rise time $\tau_{r,\,bol}$. Using
Equations 3 \& 4 of \citet{Gonzalez-Gaitan12}, which reproduce
\citet{Howell09} and use $\gamma\equiv\alpha=1.2\pm0.2$, we find
$M_{^{56}\mathrm{Ni}} = (0.44\pm0.08)\times(1.2/\alpha)\
\mathrm{M}_{\sun}$. The $\alpha$ dependency is stated explicitly due
to its large influence on the final $^{56}\mathrm{Ni}$ mass
\citep{Scalzo10}, and its uncertainty was propagated in the
calculations. If one assumes $\alpha=1.0 \pm 0.2$ (Arnett's rule) as
in \citet{Stritzinger06}, \citet{Wang09a} and \citet{Hayden10}, we
obtain $M_{^{56}\mathrm{Ni}} = 0.53\pm0.11\ {\rm M}_{\sun}$. This
value is compatible with the amount of $^{56}\mathrm{Ni}$ mass that
the explosion models used by \citet{Ropke12} were set up to produce
($\sim0.6\ \mathrm{M}_{\sun}$).

\subsection{Spectral indicators}

Several spectral indicators defined using maximum light spectral
features were measured for the SN~2011fe spectrum closest to $B$
maximum light: the $\mathcal{R}_\mathrm{Si}$ depth ratio and the
$\mathcal{R}_\mathrm{Ca}$ and $\mathcal{R}_\mathrm{SiS}$ flux ratios
\citep{Nugent95, Bongard06}; the $\mathcal{R}_{642/443}$ flux ratio
\citep{Bailey09}; the \ion{Ca}{ii} H\&K \citep{Walker10},
\ion{Si}{ii}~\wl 4131 \citep{Bronder08, Arsenijevic08, Chotard11},
\ion{Si}{ii}~\wl 5972 \& \wl 6355 \citep{Hachinger06, Branch06,
  Branch09} and \ion{C}{ii}~\wl 6580 (\cf\ Sect.~\ref{sec:carbon})
equivalent widths. The \ion{S}{ii}~\wl 5640, \ion{Si}{ii}~\wl 6355
\citep{Benetti04,Benetti05,Hachinger06} and \ion{C}{ii}~\wl 6580
feature velocities were also measured for all of the spectra for which
the features were detectable. For most of these spectral indicators
(all except $\mathcal{R}_\mathrm{SiS}$ and $\mathcal{R}_{642/443}$), a
precise estimate of the extrema wavelength and flux positions
enclosing the feature or defining the velocity is needed in order to
make a proper measurement. This was done by automatically measuring
the maximum/minimum flux of the smoothed spectra in a given wavelength
range, and confirming each measurement visually.  The uncertainties on
each of these values were derived using a Monte Carlo procedure that
takes into account the impact of the method used to select the feature
boundaries, the extrema wavelength and flux positions, as well as the
small contribution of the photon noise. A detailed description of the
measurement method with the corresponding results on a large part of
the \snf\ data set will be presented by N.~Chotard et al. (in
preparation).

\begin{figure}
  \centering
  \includegraphics[width=0.49\textwidth,clip=true]{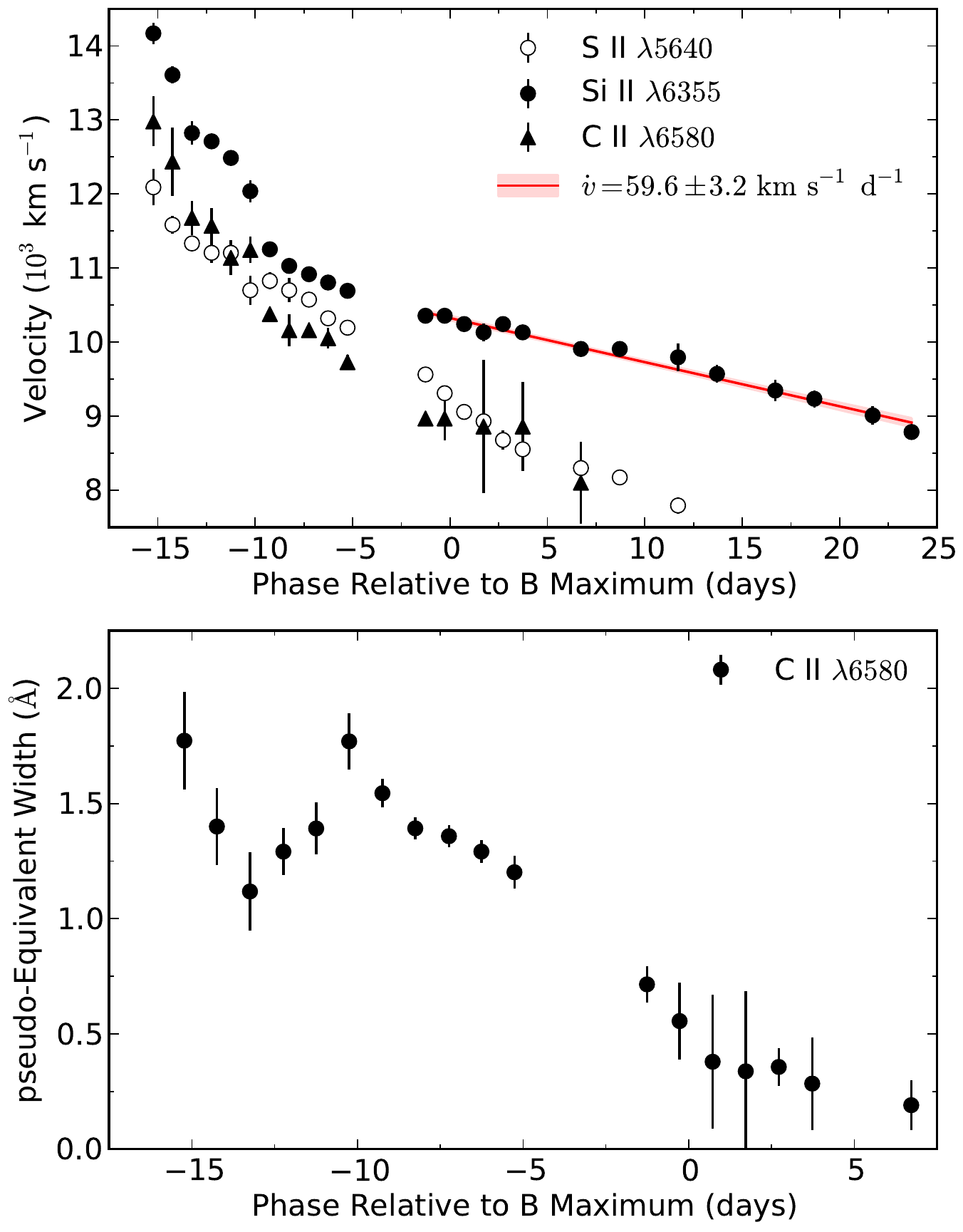}
  \caption{Phase evolution of the expansion velocities of
    \ion{S}{ii}~\wl 5640, \ion{Si}{ii}~\wl 6355, \ion{C}{ii}~\wl 6580
    (top) and the pseudo-equivalent width of \ion{C}{ii}~\wl 6580
    (bottom). A linear fit for the post-maximum velocity gradient of
    \ion{Si}{ii}~\wl 6355 is overplotted as a red line.}
  \label{fig:v_EW}
\end{figure}

The values of these measurements made on the spectrum closest to
maximum light appear in Table~\ref{tab:summary}, while the phase
evolution of the expansion velocity of \ion{S}{ii}, \ion{Si}{ii},
\ion{C}{ii} and the pseudo-equivalent width of \ion{C}{ii}~\wl 6580
are plotted in Fig.~\ref{fig:v_EW}. The velocity gradient of
\ion{Si}{ii}~\wl 6355, as defined by \citet{Benetti05}, was computed
using a linear least squares fit for $-\,2 < t < +\,25$~d overplotted
on Fig.~\ref{fig:v_EW}, and we find
$\dot{v}=59.6\pm3.2$~km\,s$^{-1}$\,d$^{-1}$. The velocity evolution of
\ion{S}{ii}~\wl 5640, which \citet{Benetti04} suggests is an effective
probe of the true photospheric velocity, is similar to that observed
by those authors for SNe 1998bu, 1994D and 1990N. It decreases from
$\sim12\,000$~km\,s$^{-1}$ at $-\,15$~d to approximately
$8\,000$~km\,s$^{-1}$ at $+\,10$~d, passing through
${\sim9\,500}$~km\,s$^{-1}$ at maximum.

\begin{figure}
  \centering
  \includegraphics[width=0.49\textwidth,clip=true]{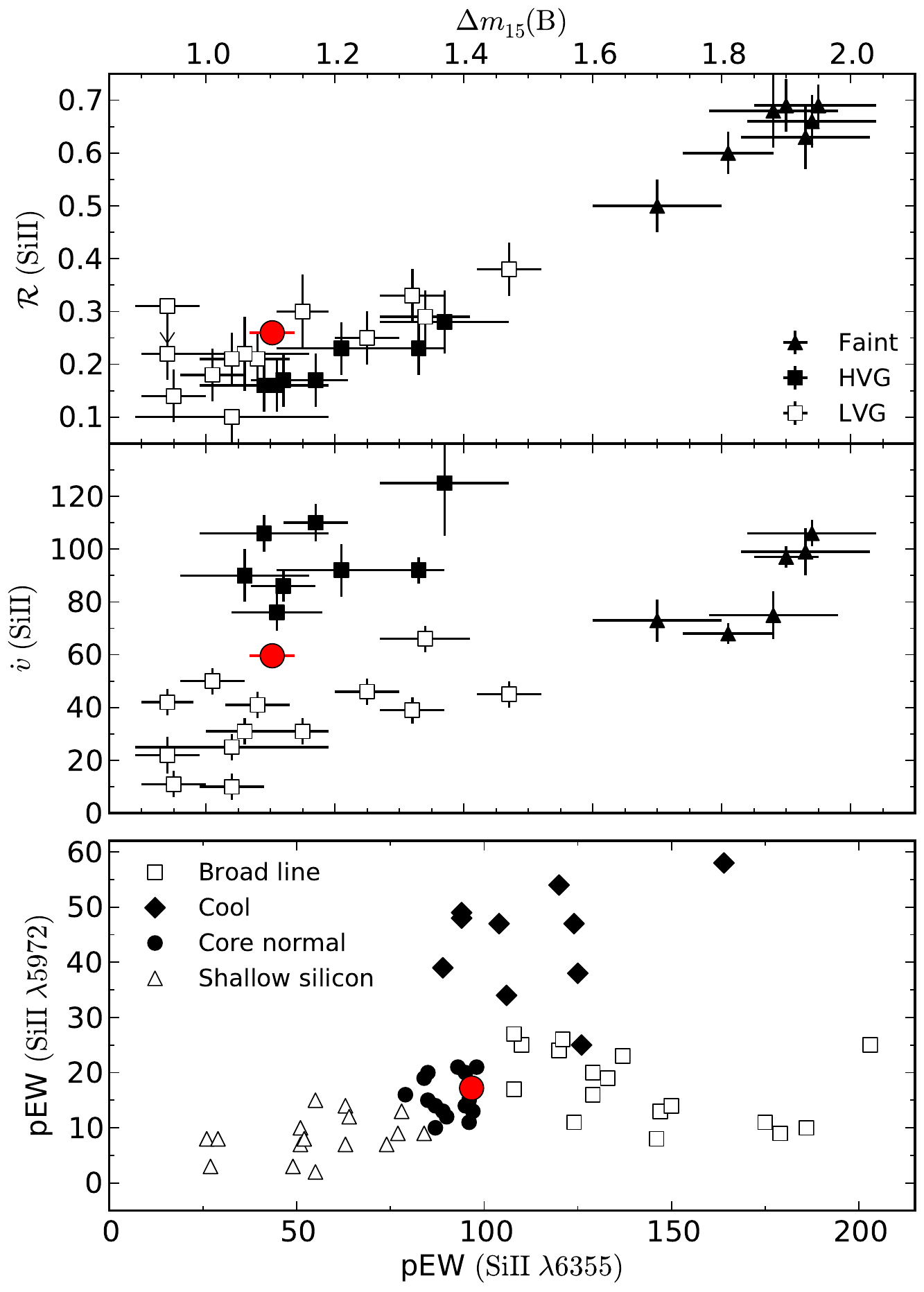}
  \caption{Top: $\mathcal{R}$(\ion{Si}{ii}) and
    $\dot{v}$(\ion{Si}{ii}) versus SN decline rate $\Delta
    m_{15}$. Comparison data and sub-classes from~\citet{Benetti05},
    updated with measurements by~\citet{Hachinger06}
    and~\citet{Taubenberger08}. Bottom: pEW (\ion{Si}{ii}~\wl 5972)
    versus pEW (\ion{Si}{ii}~\wl 6355). Comparison data and
    sub-classes from~\citet{Branch09}. SN~2011fe is represented by a
    filled red circle on all panels.}
  \label{fig:benetti_branch}
\end{figure}

Using these spectral indicators and the light curve decline rate (\cf\
Table~\ref{tab:summary}) we place SN~2011fe within the \snia\
classification schemes proposed by \citet{Benetti05} and
\citet{Branch06}, as seen in Fig.~\ref{fig:benetti_branch}. From this
comparison, SN~2011fe seems to be a spectroscopically ``core normal''
\snia\ whose expansion velocity rate of change lies close to the
separation between the ``low'' and ``high'' velocity gradient groups as
defined by
\citet[][$\dot{v}=70$~km\,s$^{-1}$\,d$^{-1}$]{Benetti05}. It is also a
``normal'' \snia\ according to the \citet{Wang09a} definition, based
on the \ion{Si}{ii}~\wl 6533 expansion velocity at maximum (\cf\ top
panel of Fig.~\ref{fig:v_EW}).

\subsection{SYNAPPS fitting}

\begin{figure*}[t]
    \centering
    \includegraphics[width=0.9\textwidth,clip=true]{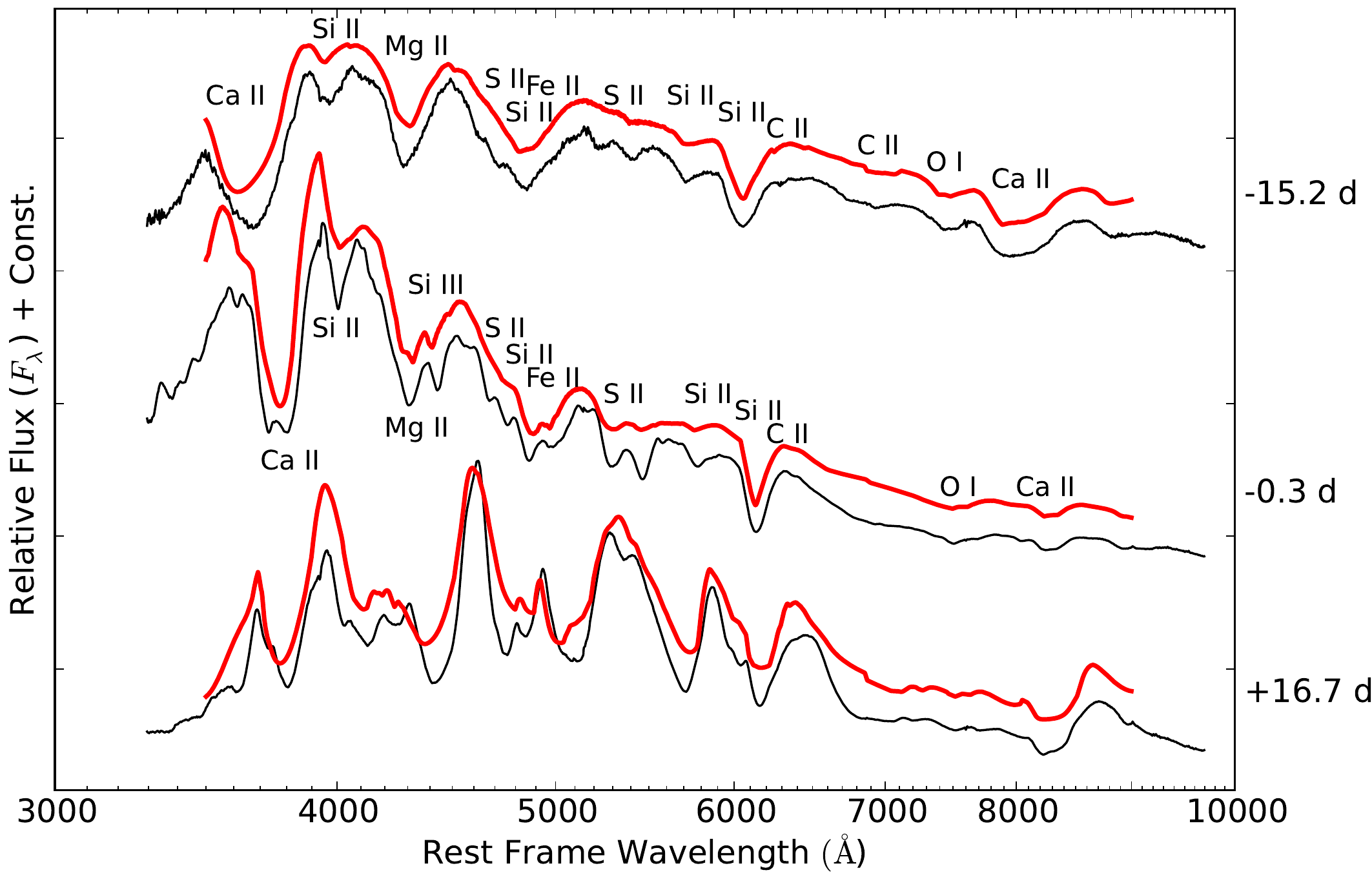}
    \caption{\synapps\ fits to selected \snifs\ spectrophotometry of
      SN~2011fe.  Major ion signatures in the first two spectra are
      shown.  The identified species are typical of a normal \snia\
      early on and near maximum light.  Two weeks after maximum, the
      spectrum is dominated by numerous blends from iron-peak elements
      (\ion{Cr}{ii}, \ion{Fe}{ii}, \ion{Co}{ii}) blueward of about
      5000~\AA.}
    \label{fig:synapps}
\end{figure*}

We analyze the spectral time series of SN~2011fe with the help of the
highly parameterized \synapps\ fitting code \citep{Thomas11a}.  The
objective is to identify ions by their spectroscopic signatures taking
line-blending into account.  The main radiative transfer assumptions
underlying \synapps\ are: spherical symmetry, a sharply-defined
blackbody-emitting pseudo-photosphere, pure-resonance line transfer
under the Sobolev approximation, with line opacity parameterized
radially according an exponential functional form and per line assuming
Boltzmann excitation.  \synapps\ combines this parameterized spectrum
synthesis calculation with a parallel non-linear optimization framework
to reduce the need for tedious interactive adjustment of fit parameters
and to assure more systematic sampling of the parameter space.  The
assumptions are simple but sufficient for our immediate purpose; a
detailed abundance tomography analysis with more sophisticated tools
will be the subject of future work.

Early, at-peak, and two weeks post-peak fits to \snifs\ data appear in
Fig.~\ref{fig:synapps}.  Extreme late-time data (after $+70$~d) are
arguably more difficult to analyze with \synapps\ as its physical
assumptions are less applicable then.  The evolution of the spectral
features follows the usual pattern of normal \sneia.  Early on, strong
contributions from \ion{O}{i}, \ion{Mg}{ii}, \ion{Si}{ii}, \ion{S}{ii}, and
\ion{Ca}{ii} are detected.  \ion{C}{ii}~\wl 6580 and \ion{C}{ii}~\wl 7234
are detected with high confidence.  High velocity components for
\ion{Si}{ii} and \ion{Ca}{ii} features are needed to achieve a good fit
in the first spectrum.  By maximum light the high velocity features have
weakened and only photospheric-velocity opacity components are needed.
The \ion{C}{ii}~\wl 6580 feature is reproduced by \synapps\ to some
extent at maximum, but the fit is not perfect.  At about two weeks
after maximum, the spectrum is dominated by lines from \ion{Fe}{ii}.  We
find evidence that some \ion{Cr}{ii} and/or \ion{Co}{ii} may be present in
the spectrum at this point, though \citet{Parrent12} did not invoke it
in their analysis.  In all other respects we confirm the analysis
presented by \citet{Parrent12}.

\subsection{Carbon signatures}\label{sec:carbon}

\begin{figure}
    \centering
    \includegraphics[width=0.45\textwidth,clip=true]{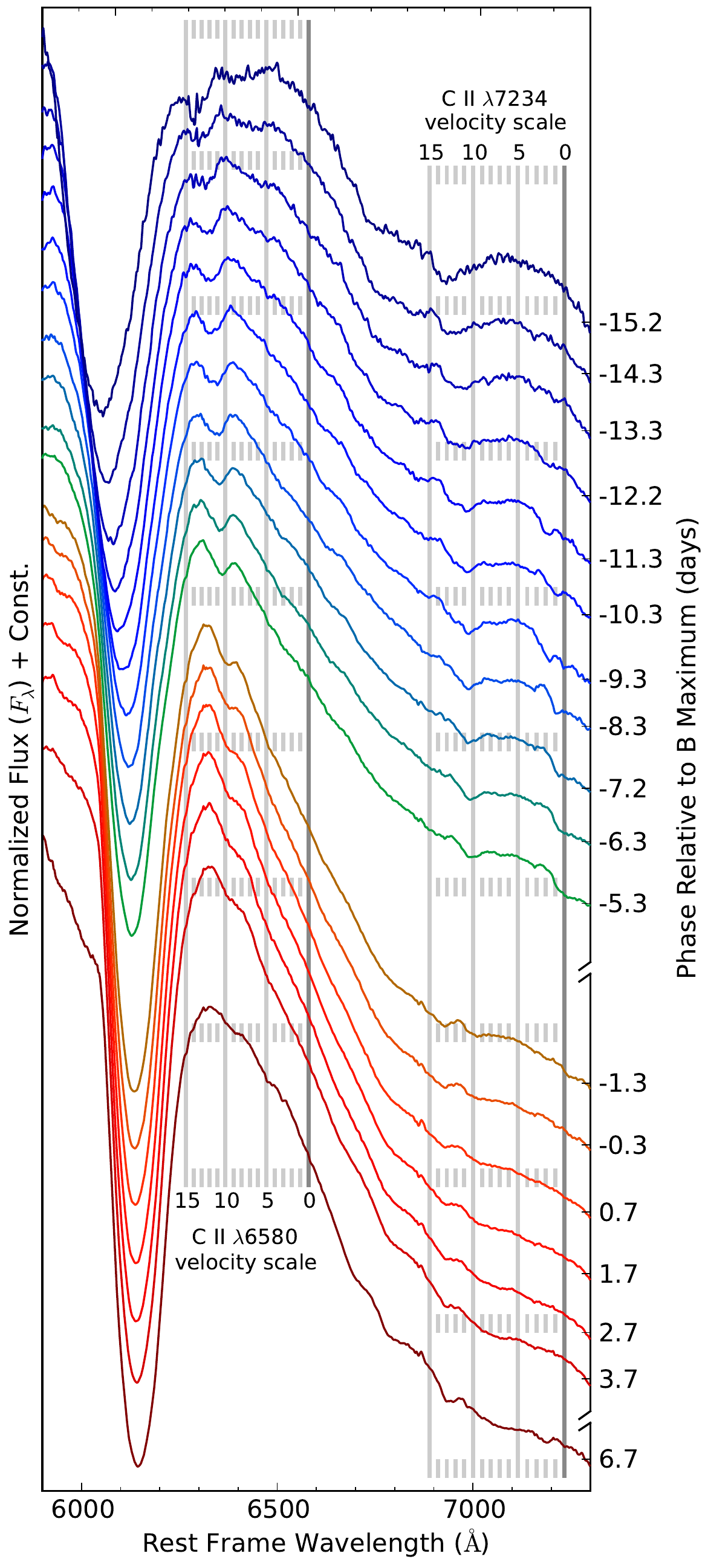}
    \caption{Evolution of \ion{C}{ii}~\wl 6580 and \wl 7234 features in
    the spectral time series of SN~2011fe.  Two axes depicting
    blueshift with respect to 6580~\AA\ and 7234~\AA\ are overlaid,
    with velocities given in units of $10^3$ km~s$^{-1}$.
    \ion{C}{ii}~\wl 6580 begins as a notch which gradually recedes in
    velocity and weakens progressively until it is difficult or
    impossible to reliably identify at $+\,6.7$~d.  \ion{C}{ii}~\wl 7234
    follows a similar evolution, but disappears by $-\,1.3$~d.}
    \label{fig:carbon}
\end{figure}

The spectra of SN~2011fe contain signatures of unburned carbon from the
earliest observations through maximum \citep{ATEL3583, Parrent12}.
Carbon in \snia\ ejecta is most robustly detected using \ion{C}{ii}~\wl
6580 absorption signatures in pre-maximum spectra \citep{Thomas07,
Parrent11, Thomas11b, Folatelli12, Silverman12, Blondin12}.  High S/N,
mostly nightly cadence, and proper treatment of telluric absorptions
allow us to trace the detailed evolution of relatively weak
\ion{C}{ii}~\wl 6580 and also \wl 7234 in the \snifs\ time series
(Fig.~\ref{fig:carbon}).  Both signatures manifest quite clearly in
the earliest spectrum, and gradually fade away.  \ion{C}{ii}~\wl 6580 is
just barely detectable in the $+3.7$~d spectrum, and absent afterward.
After $-5.3$~d, it becomes harder to clearly identify the absorption
associated with \ion{C}{ii}~\wl 7234.

In Fig.~\ref{fig:v_EW}, the blueshift at absorption minimum
is plotted.  The velocity measurements are compared to the ones extracted
from \ion{Si}{ii}~\wl 6355 and \ion{S}{ii}~\wl 5640.  At all phases,
the measured \ion{C}{ii} velocity is smaller than that measured from
the \ion{Si}{ii} feature, and tracks that of the \ion{S}{ii} line more
closely.  This offset could be sensitive to how strong the
\ion{Si}{ii} line is relative to the other two -- increasing line
opacity in expanding atmospheres (leaving other variables fixed) tends
to blue-shift absorption minima \citep[e.g.][]{Jeffery90}. The plot
suggests that unburned material is present at roughly the same
velocities as freshly synthesized intermediate mass element ejecta.
The carbon appears to extend from $15\,000$~km~s$^{-1}$ (blue absorption
edge in the first spectrum) down to $8\,000$ or $9\,000$~km~s$^{-1}$.

We extract a pseudo-equivalent width for the \ion{C}{ii}~\wl 6580
absorption feature and plot it in Fig.~\ref{fig:v_EW} as well.
\citet{Folatelli12} suggest that the pseudo-equivalent width of this
feature may increase and then decrease with time.  This behavior was
detected by \citet{Silverman12} in SN~1994D, but the peak was not
well-sampled.  The daily follow-up cadence allows us to explore the
evolution of this quantity in detail in SN~2011fe.  The equivalent width
at first decreases for two days, increases back to roughly the same
value, and then decreases again.  This behavior was strongly correlated
with the velocity evolution in SN~1994D \citep{Silverman12} but here
no such correlation is detected.


\section{Discussion}
\label{sec:discussion}

The analysis of the \snifs\ spectrophotometric time series confirms that
SN~2011fe is a ``textbook case'' \snia.  The light curve shape and color
parameters are close to those of a fiducial SALT2 \snia\
(Table~\ref{tab:summary} summarizes these and other pertinent parameters
derived for SN~2011fe in the previous sections).  The early and
near-maximum spectra exhibit typical strong low-to-intermediate-mass ion
signatures (\ion{O}{i}, \ion{Mg}{ii}, \ion{Si}{ii}, \ion{S}{ii},
\ion{Ca}{ii}) at typical \snia\ ejection velocities.  Neither the
presence of \ion{C}{ii} nor high-velocity \ion{Ca}{ii} absorption
\citep{Mazzali05} are considered particularly unusual.  As expected,
iron-peak element signatures dominate at late times as the photosphere
recedes deeper into the ejecta.  Furthermore, there is little evidence
for substantial extinction due to dust in M101 along the line of sight.

\begin{table}
  \caption{Relevant parameters for SN~2011fe derived from this work}
  \label{tab:summary}
  \centering
  \begin{tabular}{lr}
    \hline\hline
    Parameter & Value\\
    \hline
    \multicolumn{2}{c}{Photometry}\\
    \hline\\[-2ex]
    $x_{1}$ & $-0.206 \pm 0.071$ \\
    Color & $-0.066 \pm 0.021$ \\
    Stretch\:\tablefootmark{a} & $0.969 \pm 0.010$ \\
    $\Delta m_{15}\ B$\,\tablefootmark{b} & $1.103 \pm 0.035$ \\[.5ex]
    $t_\mathrm{max}\ U$ & $55813.13 \pm 0.06$ \\
    $t_\mathrm{max}\ B$ & $55814.51 \pm 0.06$ \\
    $t_\mathrm{max}\ V$ & $55816.25 \pm 0.06$ \\
    $t_\mathrm{max}\ R$ & $55816.06 \pm 0.06$ \\
    $t_\mathrm{max}\ I$ & $55812.55 \pm 0.06$ \\[.5ex]
    $U_\mathrm{max}$ / $U_{\mathrm{max}_{B}}$\tablefootmark{c} & $9.49 \pm 0.02$  / $9.52 \pm 0.02$ mag \\
    $B_\mathrm{max}$ & $9.94 \pm 0.01$ mag \\
    $V_\mathrm{max}$ / $V_{\mathrm{max}_{B}}$ & $9.98 \pm 0.02$ / $9.99 \pm 0.01$ mag \\
    $R_\mathrm{max}$ / $R_{\mathrm{max}_{B}}$ & $10.02 \pm 0.04$ / $10.04 \pm 0.03$ mag \\
    $I_\mathrm{max}$ / $I_{\mathrm{max}_{B}}$ & $10.30 \pm 0.11$ / $10.32 \pm 0.09$ mag \\[.5ex]
    M$_\mathrm{max}\ U$ & $-19.55 \pm 0.19$ mag \\
    M$_\mathrm{max}\ B$ & $-19.10 \pm 0.19$ mag \\
    M$_\mathrm{max}\ V$ & $-19.06 \pm 0.19$ mag \\
    M$_\mathrm{max}\ R$ & $-19.02 \pm 0.19$ mag \\
    M$_\mathrm{max}\ I$ & $-18.74 \pm 0.22$ mag \\[.6ex]
    $\left<\mathrm{E}(B-V)_\mathrm{host_{\,photo}}\right>$ & $0.026 \pm 0.036$ \\[.5ex]
    $\mathrm{E}(B-V)_\mathrm{host_{\,spectro}}$ & $0.014 \pm 0.003$ \\[.5ex]
    $t_{0}\ \rm bolometric$\tablefootmark{d} & $55796.81 \pm 0.13$ \\[.5ex]
    $t_{0}\ V$\tablefootmark{d} & $55796.68 \pm 0.12$ \\[.5ex]
    $\tau_{r}\ \rm bolometric$ & $16.58 \pm 0.14$ d \\[.5ex]
    $\tau_{r}\ B$ & $17.70 \pm 0.14$ d \\[.5ex]
    $L_{bol}^\mathrm{max}$ & $(1.17 \pm 0.04) \times 10^{43}$ ergs s$^{-1}$ \\[.6ex]
    $M_{^{56}\mathrm{Ni}}$ & $(0.44 \pm 0.08) \times (1.2/\alpha)\ \mathrm{M}_{\sun}$ \\
    \hline
    \multicolumn{2}{c}{Spectroscopy}\\
    \hline\\[-2ex]
    $\mathcal{R}_\mathrm{Si}$ & 0.26 \\
    $\mathcal{R}_\mathrm{Ca}$ & 1.27 \\
    $\mathcal{R}_\mathrm{SiS}$ & 1.32 \\
    $\mathcal{R}_{642/443}$ & 0.73 \\[.5ex]
    pEW (\ion{Ca}{ii} H\&K) & $97.62 \pm 11.35$ \AA \\
    pEW (\ion{Si}{ii}~\wl 4131) & $15.33 \pm 0.07$ \AA \\
    pEW (\ion{Si}{ii}~\wl 5972) & $17.21 \pm 2.19$ \AA \\
    pEW (\ion{Si}{ii}~\wl 6355) & $96.67 \pm 0.53$ \AA \\
    pEW (\ion{C}{ii}~\wl 6580) & $0.56 \pm 0.17$ \AA \\[.5ex]
    $\dot{v}$ (\ion{Si}{ii}~\wl 6355) & $59.6\pm3.2$~km\,s$^{-1}$\,d$^{-1}$\\
    \hline
  \end{tabular}
  \tablefoot{All dates are MJD. Magnitudes are on the ``standard''
    SALT2 magnitude system, allowing direct comparison with the
    literature.  Absolute magnitudes are computed using the distance
    modulus to M101 given by \citet{Shappee11}. All spectral flux ratios
    and pseudo-equivalent widths are derived at maximum light. The photon
    noise derived uncertainties for the flux ratios are below the percent
    level.\\
    \tablefoottext{a}{Derived using Equation 6 of \citet{Guy10}.}\\
    \tablefoottext{b}{Derived using the fitted SALT2 light curve model.}\\
    \tablefoottext{c}{Observed magnitude at the time of \emph{B}-band
      maximum.}\\
    \tablefoottext{d}{Derived assuming a $t^{2}$ power law.}
  }
\end{table}

The high cadence of observation, broad wavelength range, high S/N, and
good calibration make the data set presented here not only an asset for
studying \snia\ physics in detail, but also for simulations and SN
cosmology systematic error analysis.  In this section, we use our
observations of SN~2011fe to consider a few fundamental questions
about the conventional analysis of \sneia\ in those areas.  This gives
us an opportunity to demonstrate some generically useful features of the
data.

\subsection{Comparisons with other \sneia}

\begin{figure*}
   \centering
   \includegraphics[width=0.49\textwidth,clip=true]{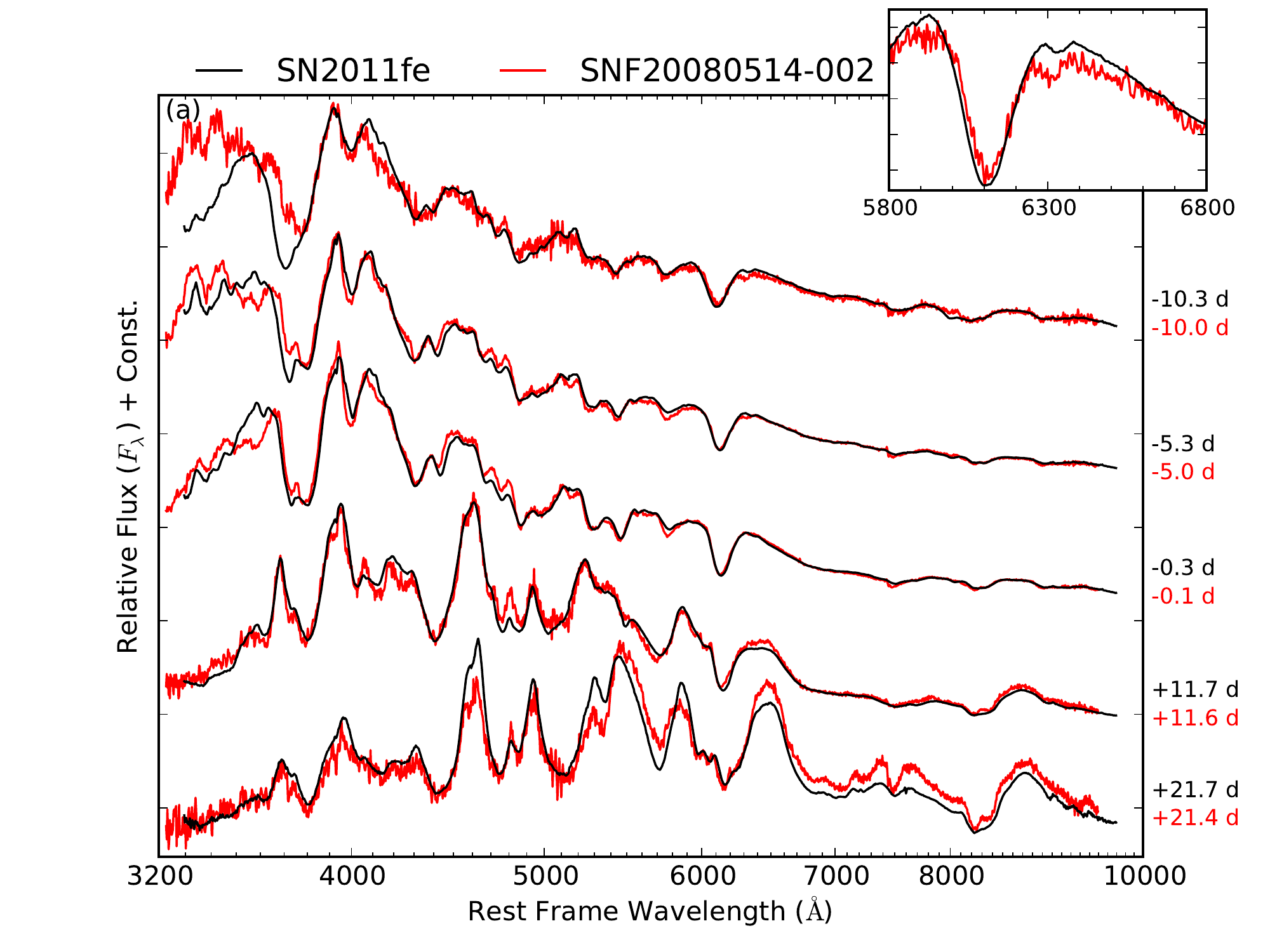}
   \includegraphics[width=0.49\textwidth,clip=true]{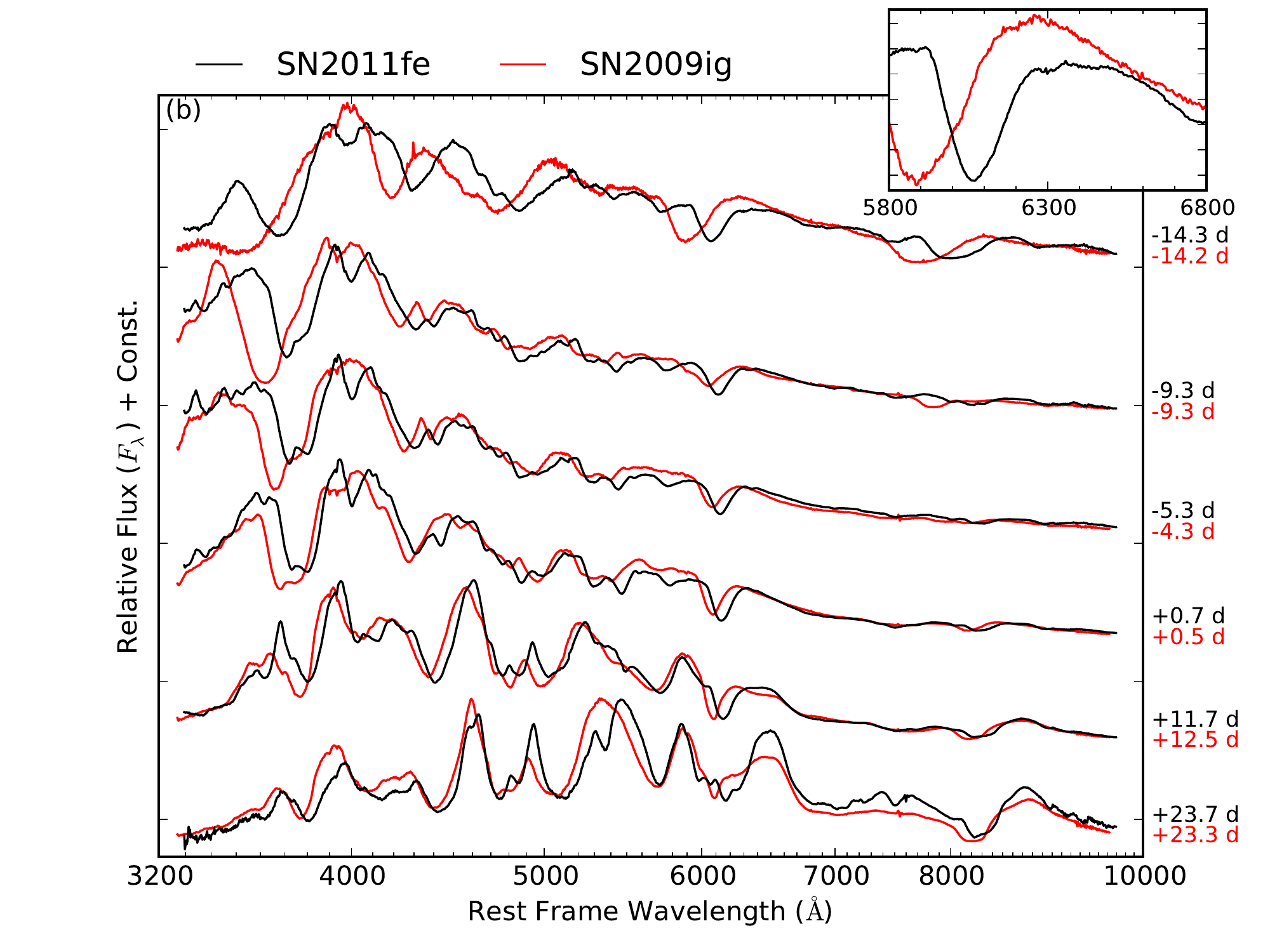}
   \caption{Comparison of selected epochs of SN~2011fe with SNF20080514-002
     (left, panel a) and SN~2009ig (right, panel b).  All observations
     were obtained with \snifs.  Insets in each panel show the region
     around the \ion{C}{ii}~\wl 6580 absorption notch (5800--6800~\AA\
     rest-frame) in the earliest pair of spectra compared.}
   \label{fig:compare}
\end{figure*}

Figure~\ref{fig:compare} compares selected \snifs\ observations of
SN~2011fe with some of SNF20080514-002 \citep{ATEL1532} and SN~2009ig
\citep{Kleiser09}.  Here we normalize the mean flux between 3500 and
7500~\AA\ in each spectrum involved, and overlay matched phases.  The
relative flux as a function of phase is not preserved, but this allows
us to compare spectral features.  These two \sneia\ are selected for
comparison because they have been identified previously as basically
``normal'' \sneia\ \citep{Thomas11b, Foley12a}.

SNF20080514-002 and SN~2011fe are a good match at all phases compared,
but they are not identical twins (Fig.~\ref{fig:compare}a).  The
agreement is best redward of 4000~\AA, in terms of what features are
present and their morphologies.  For example, both objects have strong
\ion{C}{ii}~\wl 6580 absorption notches early on (see inset, near
6300~\AA).  The double notch \ion{S}{ii} features near 4700~\AA\ and
other small-scale features are also matched well.  In contrast, however,
the $-\,10.2$~d spectrum of SN~2011fe exhibits a \emph{very} slight
high-velocity \ion{Ca}{ii} infrared triplet absorption that is weaker or
absent from the corresponding SNF20080514-002 spectrum.  This is
mirrored in the near-UV where the \ion{Ca}{ii}~H\&K absorption feature
extends to higher velocity in SN~2011fe than in SNF20080514-002.  In
fact, at this phase the spectra blueward of 3800~\AA\ are markedly
dissimilar.  Five days later, the morphologies of the near-UV spectra
appear to have converged together.  About three weeks past maximum, some
overall color difference is apparent between the two SNe, but again the
spectral features are quite similar.

There is substantial evidence in the literature for greater diversity in
the near-UV properties of \sneia\ than in the optical \citep{Ellis08,
Brown10, Milne10, Cooke11, Wang12, Foley12b}.  The near-UV behavior
depicted in Fig.~\ref{fig:compare}a supports the idea that this
extends to the time domain as well.  Increased diversity in the near-UV
at early times relative to later phases seems plausible from a physical
standpoint.  This region of the spectrum is particularly sensitive to the
composition, density, and temperature of the outer layers of the ejecta,
and at early times the spectrum forms in these layers.

SN~2009ig (Fig.~\ref{fig:compare}b) provides an excellent foil for
SN~2011fe.  \citet{Foley12a} noted the relatively high blueshifts of
absorption features in its spectra, and here we can see the contrast
with the more typical blueshifts exhibited by SN~2011fe.  In the
earliest spectrum we clearly see the systematically higher velocity
absorptions of \ion{Ca}{ii}~H\&K and infrared triplet and \ion{Si}{ii}~\wl
6355 in SN~2009ig.  The blue edges of these features weaken with time,
but overall the absorption lines remain faster in SN~2009ig than in
SN~2011fe.  An inset compares the region of the spectrum around the
\ion{C}{ii}~\wl 6580 notch in the first spectrum of either SN.  A robust
notch is seen in SN~2011fe, but any \ion{C}{ii} absorption in SN~2009ig
at the same phase is much weaker \citep{Parrent11, Foley12a}.

The above meta-comparison demonstrates a key benefit of the dense
temporal sampling of our SN~2011fe spectrophotometry, that it is a
trivial matter to find SN~2011fe spectra to compare to spectra of other
SNe at practically exactly the same phase.  This should be a highly
useful feature of the data set for other researchers who wish to
contrast their observations with a classic, normal \snia\ in a
systematic way.

\subsection{Comparisons with spectral surface templates}\label{sec:template-comp}

\begin{figure}
  \centering
  \includegraphics[width=0.49\textwidth,clip=true]{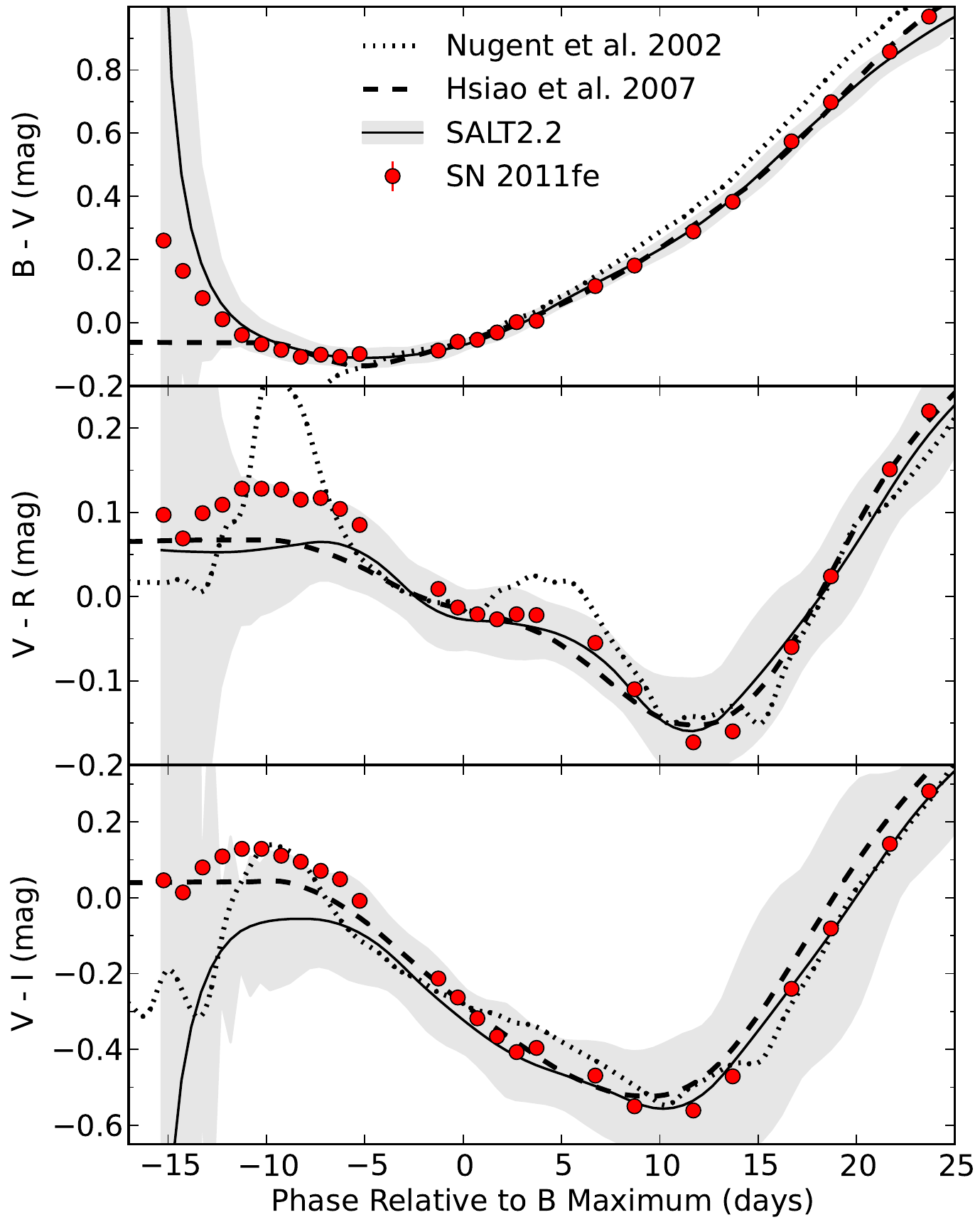}
  \caption{Synthetic color curves of SN~2011fe from \snifs\
    spectrophotometry. Both spectral templates from
    \citet{Nugent02,Hsiao07} are corrected to the observed stretch and
    peak colors. The SALT2 curves are derived from the fitted light
    curves on the different bands. The error bars on the SN~2011fe
    values are smaller than the plot symbols.}
  \label{fig:colorcurves}
\end{figure}

Here we examine how well the color and detailed spectral evolution of
SN~2011fe are described by spectral surface ($\lambda,t$) templates
used for conventional light curve fitting and \emph{K}- and
\emph{S}-corrections.  Such templates should do well at describing
``archetypical'' \sneia.  Our default expectation is that the template
spectra should have high fidelity to the ``ground truth'' time series,
that of SN~2011fe.  The comparison allows us to gauge the level of
that fidelity, and identify transient or systematic deficiencies in
the templates themselves.  We begin by examining the agreement between
true and template-synthesized color curves and then take a closer look
at the agreement at the fine spectral level.

$B-V$, $V-R$, and $V-I$ color curves of SN~2011fe, synthesized from
\snifs\ spectrophotometry (Sect.~\ref{sec:time-series}) up to 40 days
past explosion, using BM12 passbands and zero points, appear in
Fig.~\ref{fig:colorcurves}.  Color curves synthesized from standard
templates \citep{Nugent02, Hsiao07}, corrected to the stretch and
observed colors at peak luminosity, are shown for comparison.

The observed $B-V$ color curve closely follows the Hsiao template
starting at about $-\,10$~d, while the match with the Nugent template is
only good after about $-\,5$~d and diverges slightly starting at $+\,5$~d.
For very early epochs (before $-\,10$~d) neither template seems to be
a good match.

In the case of the $V-R$ color evolution, SN~2011fe matches reasonably
well with the Hsiao template after maximum light, while earlier phases
differ significantly.  Interestingly, the Nugent template shows a large
deviation from the Hsiao template at the same phases that the Hsiao
template differs from the SN~2011fe observations; notably the ``bumps''
at $-\,10$~d and (to a lesser extent) $+\,5$~d. While this difference is not
as pronounced as the one between both templates, one could argue that in
the small sample of \sneia\ used to construct the \citet{Nugent02}
template, there are spectra whose time evolution is closer to that of
SN~2011fe than the average \snia\ spectral time series represented by
the \citet{Hsiao07} template. This appears to be confirmed by the fact
that the two \sneia\ that SN~2011fe is most similar to spectroscopically
\citep[SN~1992A and SN~1994D, according to][]{Nugent11} are those
contributing the most spectra to the Nugent template, along with SN~1989B.

The evolution of $V-I$ is similar to that of $V-R$.  After maximum
light, the data agree rather well with the Hsiao template, with the
Nugent template being a better match from $+\,15$~d onwards. At
earlier phases the observations display a significant systematic
departure from any of the templates, being closer to the Nugent
template at about $-\,10$~d. From both this color curve and the
previous, it is apparent that the second \snifs\ spectrum (MJD 55800)
has a problem at redder wavelengths, being fainter than we would
expect it to be. This systematic error ($\sim-0.03$ and $\sim-0.04~$mag
for \emph{RI} respectively) is close to the statistical uncertainty
for the flux calibration of that night and seems to be due to
extraction problems, not with the SN itself but rather with the sole
standard star used for calibration.

The ``fitted'' SALT2 color curves and error bands are also plotted in
Fig.~\ref{fig:colorcurves}.  For these, we can see that the fitted
colors are accurate throughout all of the phases after $-\,5$~d, but
start to break away from the observations at phases $\la-\,10$~d for
$B-V$ and before $-\,5$~d for both color curves using the redder parts
of the spectra. For the latter, the fits are systematically too red,
though the observations are within the quoted model errors.  This is
not unexpected since the SALT2 model is trained on both nearby and
high-z \sneia\ observations, with the former often lacking
observations at very early phases, and the latter being affected by
low signal-to-noise for the redder bands due to quantum efficiency
drops and atmospheric extinction effects.

\begin{figure*}
  \centering
  \includegraphics[width=0.9\textwidth,clip=true]{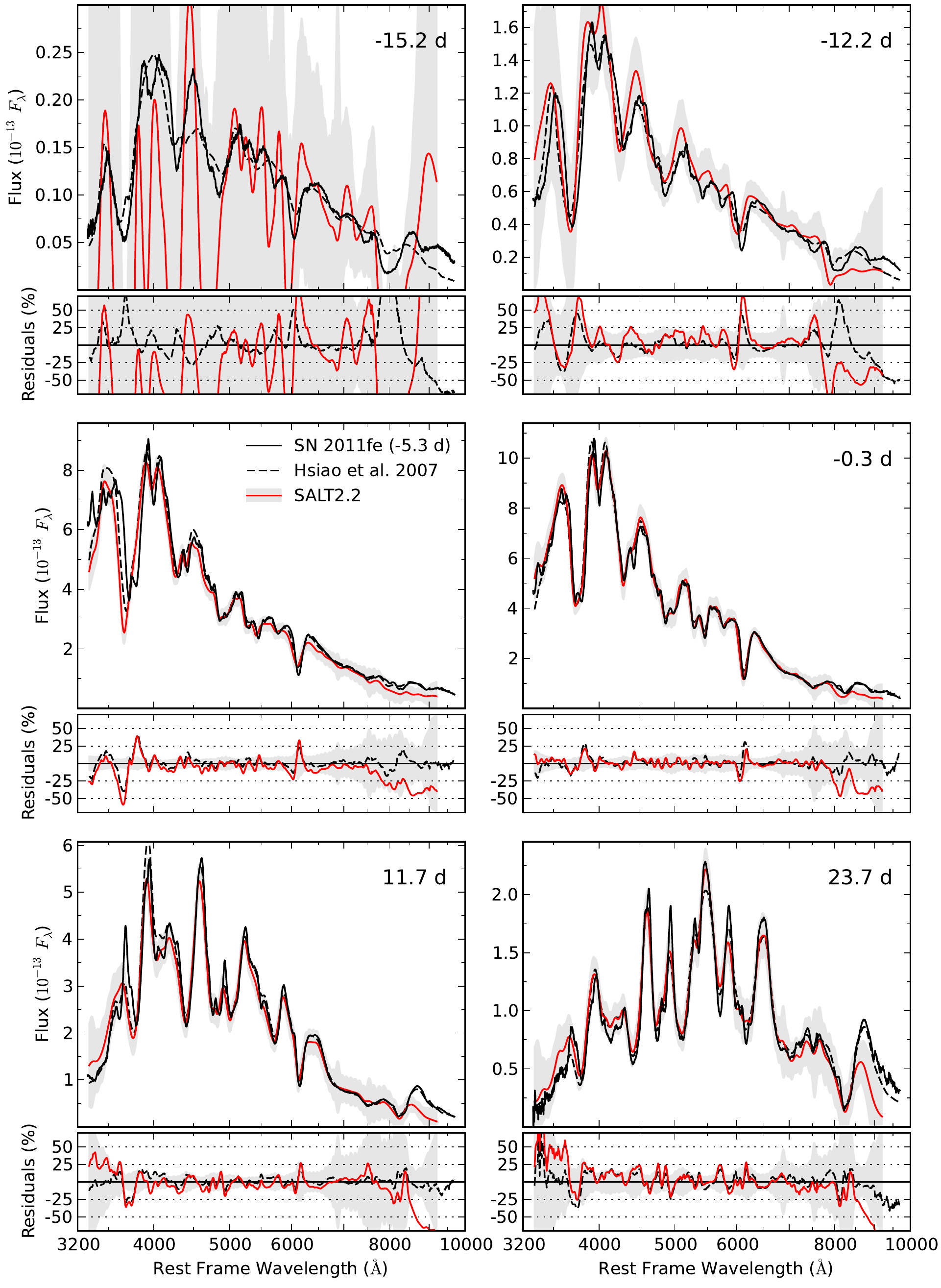}
  \caption{Comparison of six different phases of SN~2011fe (black
    line) with the \citet{Hsiao07} (black dashed lines) and SALT2 (red
    line with grey error model) \sneia\ spectral surface
    templates. The \citet{Hsiao07} spectra are warped to match the
    observed broad-band colors of SN~2011fe. SALT2 spectra are created
    directly from the template using the light curve fit
    parameters. The lower panels plot the percent residuals of each
    template with relation to the observed SN~2011fe spectra.}
  \label{fig:templates}
\end{figure*}

Next we consider agreement at the fine spectral level.
Fig.~\ref{fig:templates} is a comparison of six representative \snifs\
observations of SN~2011fe with two \snia\ spectral surface templates
available in the literature, that from \citet{Hsiao07} and SALT2.  The
rapidly changing spectral features during the rise in luminosity to peak
are the most interesting to consider, so four spectra leading up to
maximum are shown.  Observations after peak brightness are represented
by the spectra at $+\,11.7$ and $+\,23.7$~d.

The Hsiao template spectra were warped as in \citet{Foley12a}, in
order to emulate their usage in \emph{K}- or \emph{S}-correction
computations.  For each phase the template spectrum is normalized by a
cubic spline fitted to the five ratios of synthetic \emph{UBVRI}$_{\rm
  SNf}$ fluxes between the template and our spectrum for that same
phase. In this way the template is forced to match the observed
broad-band colors of SN~2011fe.

The SALT2 spectra, on the other hand, are created directly from the
template using the light curve fit parameters ($x_{0}$, $x_{1}$ and $c$)
found in Sect.~\ref{sec:salt2-fit}, without any additional warping over
wavelength or time.  Thus, the projected spectrum from SALT2 is what it
predicts the underlying SN spectrum looks like without modification.  No
scaling in flux is added to improve the agreement, even cosmetically.
The SALT2 model $1 \sigma$ error is depicted as the gray shaded region
in each panel of Fig.~\ref{fig:templates}.  The lower plot in each
panel shows the percent residual differences between each template
and the observations.

Though there are both transient and persistent artifacts in the
residuals, a general trend is also evident.  The agreement between the
templates and SN~2011fe improves as maximum light is approached.  Near
maximum, where the population of \sneia\ available for constructing
either template is very well sampled, the discrepancies are relatively
smaller.  After this point, glitches in the residuals become larger
again.  We note that the residuals are decidedly not the result of a
bias induced by comparing the spectra to templates scaled by synthetic
photometry, the high S/N of the observations limits this effect to less
than 1\%.

We start by examining the agreement between SN~2011fe and both templates
at phases $t<-10$~d.  The disagreement between the projected SALT2
spectrum in Fig.~\ref{fig:templates} and SN~2011fe at $-15.2$~d is at
first striking.  The SALT2 error model strongly deweights this
observation (and others at the earliest phases, see
Sect.~\ref{sec:salt2-fit}), so the mismatch is not ultimately
catastrophic.  To an extent, this suggests that these early epochs are
underutilized in modeling \sneia\ for cosmological applications today.
By $-12.2$~d, the template shows signs of converging toward SN~2011fe
but both the estimated errors and residuals remain large.  The Hsiao
template is not a particularly good match to SN~2011fe at these phases
either.  In particular, it seems to systematically underestimate the
depths and widths of the stronger absorption features.  The template is
based on a comparatively small number of spectra at early phases: 13
spectra with phases between $-15$ and $-10$~d, compared to 50 spectra or
more in each 5-day bin from $-10$ to $+15$~d.  The template is obviously
much more susceptible to biases in the input sample before $-10$~d.

SALT2 systematically under-estimates the flux at wavelengths redward of
about 7500~\AA, at all phases.  Setting aside the case of $-15.2$~d, the
SALT2-predicted flux at these wavelengths always appears to be smaller
than the true value, and the shape of the broad \ion{Ca}{ii} infrared
triplet feature is poorly reconstructed.  The warped Hsiao
template seems to have higher fidelity to SN~2011fe at these wavelengths
than SALT2 in all but the first two spectra shown. While there are
similarly sized offsets at the bluest wavelengths, they are not either
systematically low or high as at the red end.

Interestingly, at all phases where the bellwether \ion{Si}{ii}~\wl 6355
absorption is unblended and easily discernable in the observed
spectrum, it corresponds to a prominent, isolated, coherent feature of
the residuals.  The same applies to the \ion{Ca}{ii}~H\&K feature.
Discrepancies in feature strength, width, and blueshift are all to
blame. Even at maximum light, the residuals relative to SALT2 are
larger than the model error in certain places (by factors of several at
some wavelengths) while the integral of those residuals in the
photometric band considered remains small. This emphasizes the
difficulty in extracting accurate spectral templates by combining
spectra with photometry, and the usefulness of spectrophotometric SN
follow-up for this purpose. In the last spectrum, we see that most of
the features are reconstructed at the right wavelength, but the emission
profiles in either template are too low or too high.

Considering the SALT2 residuals, it at first seems plausible that
discrepancies such as those noted here are symptoms of incompleteness
in the sample used to create the template.  Merely expanding the sample
would then ameliorate any resulting errors.  This suggestion is
certainly applicable to the earliest observations, but seems less
reasonable closer to maximum light.  The fact that the SALT2 template
spectrum at $-0.3$~d very closely resembles that of a normal \snia\
suggests that quantity and quality of the underlying sample are not the
only issues.  Rather, it seems plausible that additional information in
the form of one or more model parameters (beyond light curve shape and
color) are needed to null out coherent residuals associated with major
spectral features, in particular those originating in the \ion{Ca}{ii}
H\&K and \ion{Si}{ii}~\wl 6355 features \citep[see also][]{Chotard11}.
It seems quite plausible that in these features the residuals arise from
velocity mismatch, and that if spectra were used to constrain the
template fit, the contribution to the photometric correction error could
be reduced.  Of course, spectrophotometric measurements eliminate the
problem entirely.

It is important to acknowledge that comparing templates designed to
stand in for archetypical \sneia\ to SN~2011fe may suggest an optimistic
outlook in the context of light curve fit systematics.  On the other
hand, it provides a near best-case or at least typical scenario.  To
contrast, we refer the reader to a similar comparison of SN~2009ig and
the \citet{Hsiao07} template by \citet{Foley12a}.  In that particular
case, the discrepancies in spectral evolution presented were portrayed
as severe.  We note that we performed the same projected SALT2 residuals
analysis as depicted in Fig.~\ref{fig:templates} for SN~2009ig, and
find the same kinds of residual structures with often larger but at
times smaller magnitude than is seen in SN~2011fe.  A more systematic
analysis of spectral surface templates is underway and will be presented
elsewhere (Saunders et~al., in preparation).

\subsection{Detecting \ion{C}{ii} in \snia\ spectra}

Since \sneia\ are likely the thermonuclear incineration of carbon-oxygen
white dwarfs, estimates of the total mass and spatial distribution of
unprocessed carbon in SN ejecta provide a way to constrain the explosion
mechanism.  Pure turbulent deflagration models predict that a large
amount of unprocessed carbon may remain after explosion
\citep[e.g.,][]{Gamezo03, Ropke05, Ropke07}.  Models where this
deflagration is followed by a detonation phase consume much of the
remaining carbon \citep[e.g.,][]{Hoflich02, Kasen09}.  Oxygen signatures
in the photospheric phase are less useful, because oxygen is both a fuel
and product in the explosion.

The high quality of the \snifs\ time series and the clear presence of
\ion{C}{ii}~\wl 6580 provide us with the opportunity to explore some
questions about the efficiency of carbon detection in \snia\ spectra.
SN~2011fe gives us a ``ground truth'' \ion{C}{ii}~\wl 6580 detection at
high S/N needed to simulate the impact of noise and spectrograph
resolution on this principal carbon indicator.  The feature manifests as
a small notch or small-amplitude depression in flux, so both noise and
resolution are important factors in its (positive) detection.
\citet{Parrent11} briefly explore the effect of noise, but only on a
single illustrative example spectrum.  Here we perform simulations at
several phases, especially later than $-\,15$~d which is more typical for
initial \snia\ follow-up.  Specifically, we are interested in simulating
quick analyses done by observers attempting to assess whether
\ion{C}{ii}~\wl 6580 is present in freshly obtained spectra, as is
commonly described in astronomical circulars.

\begin{figure*}
    \centering
    \includegraphics[width=0.82\textwidth,clip=true]{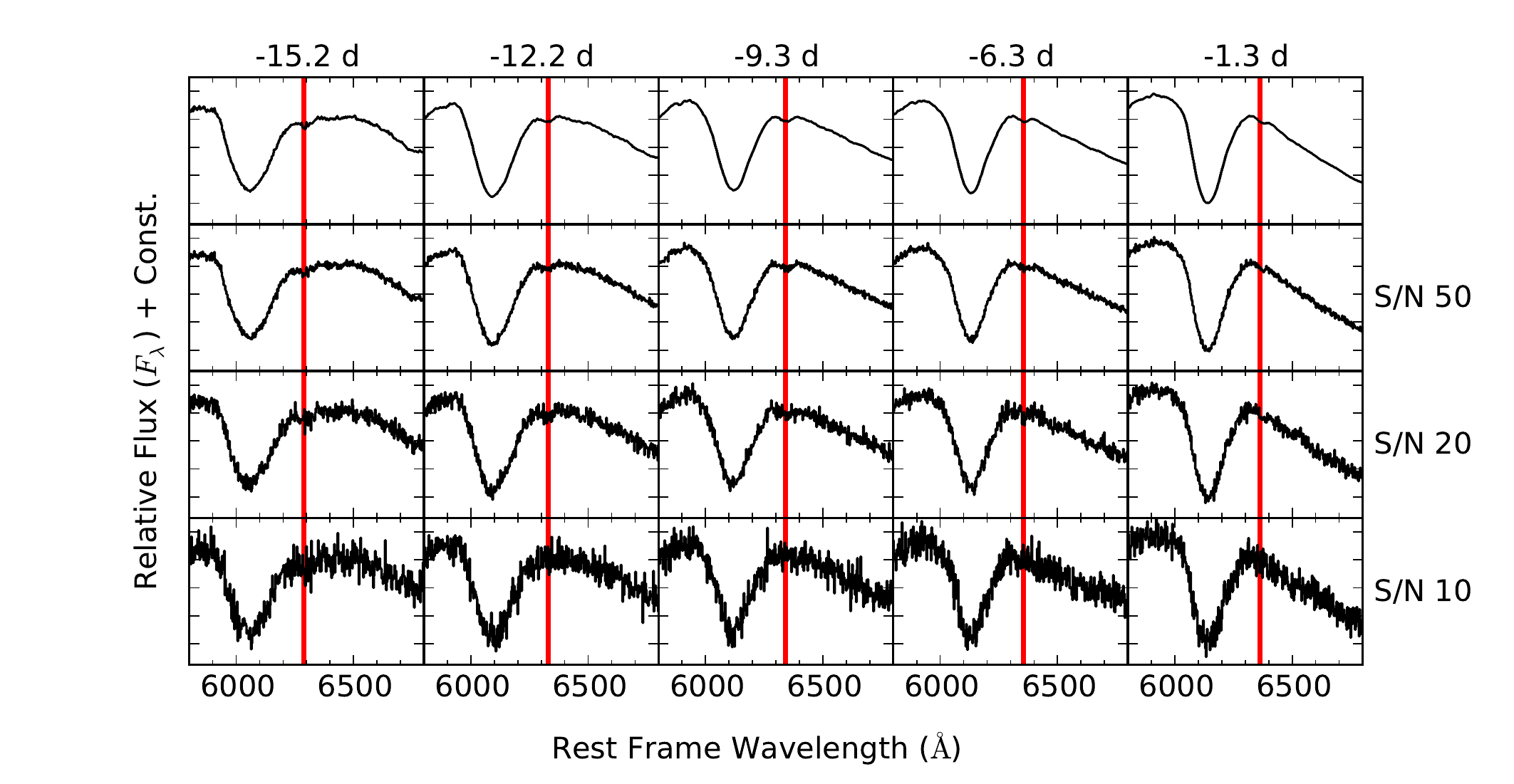}
    \caption{Effect of noise at various phases on the \protect\ion{C}{ii}~\wl
    6580 absorption notch.  The top row depicts the neighborhood of the
    feature as observed by \snifs.  Each successive row simulates
    progressively noisier observations, with the S/N per \AA\ labeled at
    right.  The red vertical line depicts the center of the \protect\ion{C}{ii}
    notch to guide the eye.  Early observations are no guarantee of a
    \protect\ion{C}{ii} detection even if it is there at lower S/N but generally
    the feature becomes harder to discern with time as the notch
    amplitude decreases.  Wavelength and flux scales depicted are the
    same in each panel.}
    \label{fig:fake_snifs}
\end{figure*}

\begin{figure*}
    \centering
    \includegraphics[width=0.82\textwidth,clip=true]{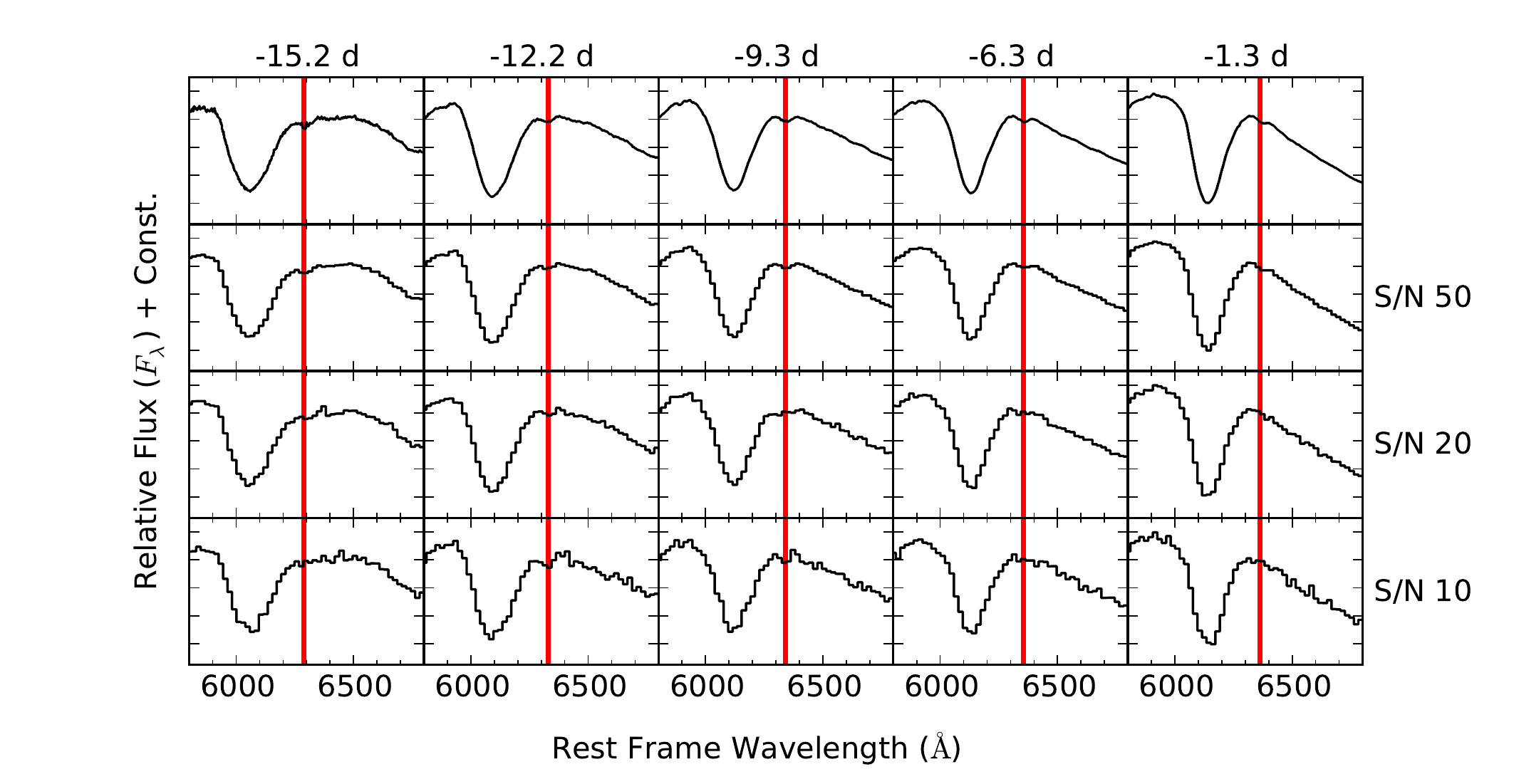}
    \caption{Same as Fig.~\ref{fig:fake_snifs} but rebinned by a
    factor of eight (bin size $\sim 19$~\AA).  Rebinning noisy data may
    improve confidence in \ion{C}{ii}~\wl 6580 detection in some cases.}
    \label{fig:fake_n8}
\end{figure*}

\begin{figure*}
    \centering
    \includegraphics[width=0.82\textwidth,clip=true]{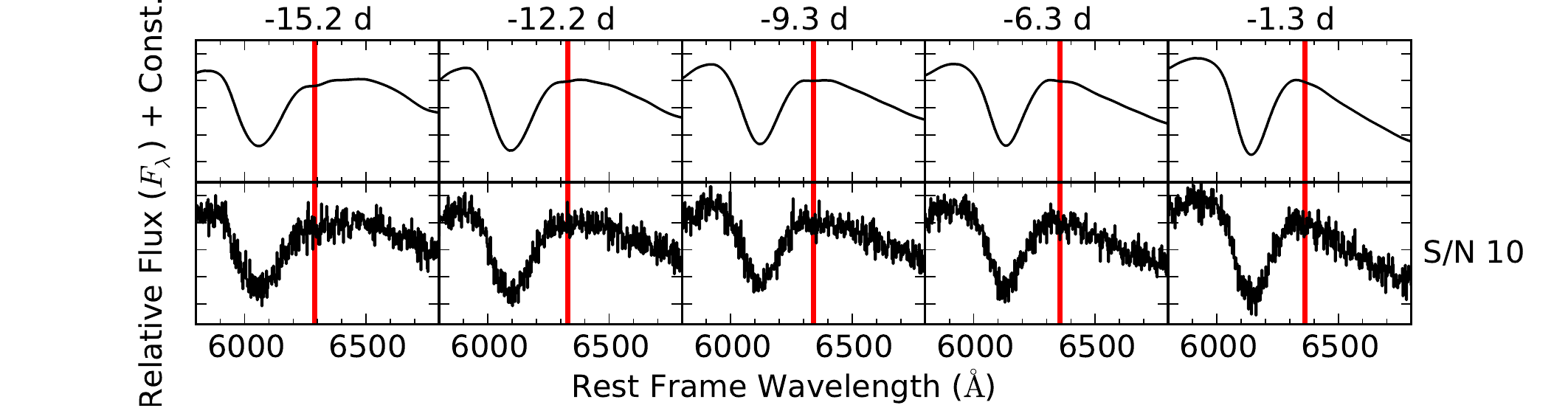}
    \caption{Same as Fig.~\ref{fig:fake_snifs} but with the spectral
    resolution degraded to $R \sim 100$.  The top row depicts the
    smoothed input to the simulation, which nearly completely destroys
    the \ion{C}{ii} signature at all epochs.  At the earliest phase a
    ``break'' in the shape of the \ion{Si}{ii}~\wl 6355 feature can be
    seen near 6250~\AA.  The break signature may be a way to detect
    \ion{C}{ii} at lower spectral resolutions and binning than our
    observations here, but only at the earliest phases.}
    \label{fig:fake_r100}
\end{figure*}

Figure~\ref{fig:fake_snifs} demonstrates the effect of lowering the S/N
in five pre-maximum spectra of SN~2011fe where \ion{C}{ii}~\wl 6580 is
clearly evident in the \snifs\ data.  The real observations of the
\ion{C}{ii}~\wl 6580 region, adopted as effectively perfect, appear in
the top row of panels in the figure.  The same spectrum is shown on each
subsequent row with gradually decreasing S/N (given as S/N per
rest-frame \AA\ in the neighborhood of the carbon notch).  At a S/N of
10 per rest-frame \AA\ (bottom row), the \ion{C}{ii} notch is completely
obscured at the native \snifs\ extraction binning (2.38~\AA).  Overall, the figure
suggests that relatively low S/N may be tolerable at early times for
making a correct carbon detection, because the feature's signal is
strongest then.  Positive identification becomes more difficult as the
feature fades away.  Indeed, it is hard to make a case that the
\ion{C}{ii}~\wl 6580 notch is clearly present at $-1.3$~d at \emph{any}
simulated S/N.

Binning the data can recover the feature, but at the cost of effectively
smoothing the data, as shown in Fig.~\ref{fig:fake_n8}.  Simulations
generated as in Fig.~\ref{fig:fake_snifs} are binned by a factor of
eight.  This is particularly detrimental at the earliest phases and,
somewhat surprisingly, at the lowest S/N ratios.  The small wavelength
scale of the feature means that it is susceptible to contrast losses if
the data are simply binned by too large a factor.  We suggest that
binning may help improve confidence in more marginal \ion{C}{ii}
detections at low S/N, but not always.  Also, systematic spectral
feature modeling \citep[e.g.,][]{Thomas11b} may eventually provide a
quantitative approach.

Finally, Fig.~\ref{fig:fake_r100} depicts simulated data as before,
but with the resolution degraded to $R \sim 100$.  The top row again
shows the \snifs\ spectra without any noise added, but with a Gaussian
filter applied.  Only a single noisy realization is plotted here, since
it is clear from the top row of the figure that at this resolution the
\ion{C}{ii}~\wl 6580 feature is obliterated nearly completely at all
phases.  Unless careful modeling is used, it is doubtful that a positive
carbon detection would be made at any phase.  A characteristic ``break''
at about 6250~\AA\ at the earliest phase in the simulated spectrum may
be a useful signature, but could be indistinguishable from normal
variation in the morphology of \ion{Si}{ii}~\wl 6355 features.

The above reinforces the somewhat obvious fact that constructing an
accurate descriptive census of carbon in \snia\ ejecta requires both
high S/N and moderately high resolution.  Earlier is better than
later, but blue-shifted \ion{C}{ii}~\wl 6580 notches ``blend'' more
readily with the \ion{Si}{ii}~\wl 6355 absorption at lower S/N.  Lower
resolution robotic integral field spectrographs, such as the SED
Machine \citep{Ben-Ami12}, are an excellent way to provide prompt
broad classifications and phase estimates for SNe, but are unlikely to
directly contribute to a carbon census in \sneia.  This will fall to
higher resolution spectrographs with larger apertures that can trigger
on early \snia\ alerts from first-response screening that such
instruments as the SED Machine provide.  Real-time forward modeling of
features that includes instrumental effects may be critical to make
the most of such data.

Given the difficulties inherent in translating early discovery into
early spectroscopic follow-up of \sneia, it seems quite possible that
photospheric-velocity carbon signatures at the earliest phases may be
anything from common to ubiquitous.  Considering also that \ion{C}{ii}
line strengths vary from object to object and as a function of phase
\citep[e.g.,][]{Parrent11, Thomas11b}, the clearest route to a proper
census is also the most challenging: orchestrating discovery and
follow-up on the same or next night for day-old \sneia.  This remains
far from routine for wide-field surveys (such as PTF).  High-cadence
surveys targeting nearby galaxies \citep[e.g., the Lick Observatory
Supernova Search,][]{Filippenko01} may slowly build up a sample of
bright objects.  For now, it seems wise to consider the current rate
estimates from the literature \citep{Parrent11, Thomas11b, Folatelli12,
Silverman12, Blondin12} to be lower limits.

The qualitative study presented here highlights the usefulness of high
S/N, flux calibrated time series to study detection systematics of
narrow and weak spectral features. Our data are a natural input for
systematics study simulations where the accurate shape of the SN
pseudo-continuum is important. Using any of the current templates for
the same purpose will yield biassed results, since they are by nature
capturing an average behavior, tuned to be useful in a different
context.


\section{Conclusion}\label{sec:conclusion}

We have presented a new atlas of spectrophotometry of SN~2011fe, a
photometrically and spectroscopically normal \snia.  SN~2011fe is not
the typical \snf\ target, and hence its calibration route is not the
standard one used for the bulk of the \snf\ data set.  Difficulties
encountered were the result of short exposure times and very high
airmass.  Still, the attained level of calibration, high S/N, and
observing cadence make this data set extremely useful for a variety of
science applications.

The SALT2 fit performed on synthesized light curves shows SN~2011fe to
have attained a $B$-band maximum of $9.94\pm0.01$~mag on MJD
$55814.51\pm0.06$, with ``standard'' lightcurve shape parameters ($x_1
= -0.206 \pm 0.071, c = -0.066 \pm 0.021$). The residual scatter when
compared with published SN~2011fe photometry is comparable to the
errors estimated for the flux calibration, and of the same order as
the SALT2 fit residuals and the scatter between independent
photometric followup campaigns. Reddening due to dust in the host
galaxy is found to be very moderate and in accordance with independent
spectroscopic determinations. An ultraviolet + optical + near-infrared
template was built from \snifs\ data and public UV spectroscopy and
NIR photometry, to construct a bolometric light curve. From it we
derive a date of explosion $t_{0}=55796.62\pm0.13$ in the assumption
of a $t^{2}$ luminosity evolution for the very early phases, with a
rise-time $\tau_{r}=16.58\pm0.14$~d and an inferred $^{56}\mathrm{Ni}$
mass of $(0.44 \pm 0.08) \times \ (1.2/\alpha)\ \mathrm{M}_{\sun}$.

The analysis of spectral indicators shows SN~2011fe to be a
spectroscopically core-normal \snia, on the lower side of the
separation between the HVG and LVG groups as defined by the rate of
change of the expansion velocity of \ion{Si}{ii}~\wl 6355
($59.6\pm3.2$~km\,s$^{-1}$\,d$^{-1}$).  The evolution of spectral
features is typical of a normal \snia, with low to intermediate-mass
elements dominating the early spectra, and iron-peak signatures
strengthening after maximum light.  High-velocity \ion{Ca}{ii} and
\ion{Si}{ii} are needed to explain the early spectra, but these weaken
by maximum light.  \ion{C}{ii}~\wl 6580 and \ion{C}{ii}~\wl 7234 are
detected in pre-maximum spectra, and our high S/N observations and
daily cadence allow us to follow the evolution of these features very
closely as they fade: apparently unburned carbon extends as deep as
$8\,000$ km~s$^{-1}$ in some normal \sneia.

The \snifs\ time series possesses a number of features that should make
it highly useful for interpreting other \sneia, investigating systematic
errors in traditional light curve analysis, and making forecasts for new
instruments' sensitivities to detecting physically important features.


\begin{acknowledgements}

  We thank the following University of Hawaii astronomers, who
  graciously granted us interrupt time so that we could observe
  SN~2011fe during its earliest phases: Colin Aspin, Eric Gaidos,
  Andrew Mann, Marco Micheli, Timm Riesen, Sarah Sonnett, and David
  Tholen.  The engineering and technical staff of the University of
  Hawaii 2.2~m telescope helped make this work possible.  We recognize
  the significant cultural role of Mauna Kea within the indigenous
  Hawaiian community, and we appreciate the opportunity to conduct
  observations from this revered site.  We thank Peter Brown and
  Ulisse Munari for providing early access to the full tables of
  \emph{Swift}/UVOT and ANS measurements, Xiaofeng Wang for providing
  the HST spectra of SN~2005cf, and Dan Birchall for his assistance in
  collecting data with \snifs.  This work was supported by the
  Director, Office of Science, Office of High Energy Physics, of the
  U.S.\ Department of Energy under Contract No.~DE-AC02-05CH11231; by
  a grant from the Gordon \& Betty Moore Foundation; in France by
  support from CNRS/IN2P3, CNRS/INSU, and PNCG; and in Germany by the
  DFG through TRR33 ``The Dark Universe.''  Some results were obtained
  using resources and support from the National Energy Research
  Scientific Computing Center, supported by the Director, Office of
  Science, Office of Advanced Scientific Computing Research, of the
  U.S.\ Department of Energy under Contract No.~DE-AC02-05CH11231.

\end{acknowledgements}

\bibliographystyle{aa}
\bibliography{bibliography}

\Online
\begin{appendix}
  \begin{figure*}
    (placeholder for animation)
    \caption{Animation of the spectral evolution of SN~2011fe for
      phases $-\,15 < t < +\,45$~d. The \snifs\ spectrophotometric
      data was interpolated linearly per wavelength bin between each
      observation date. The inset shows the light curves synthesized
      using the \emph{UBVRI}$_\mathrm{SNf}$ filter set, represented as
      colored shaded regions.}
  \end{figure*}
\end{appendix}

\end{document}